\documentclass{jfm}

\usepackage{mathpazo}
\usepackage{amsmath}
\usepackage{ucs}
\usepackage{array}
\usepackage{ifthen}
\usepackage[applemac]{inputenc}

\usepackage{pdfpages}
\usepackage[normal]{caption}
\usepackage{verbatim}
\usepackage{multicol}
\usepackage{multirow}
\usepackage{wrapfig}
\usepackage{float}
\usepackage[T1]{fontenc}
\usepackage{graphicx}
\usepackage{subfigure}
\usepackage{verbatim} 
\usepackage{upgreek}
\usepackage{wasysym}
\usepackage{numprint}
\usepackage{mathrsfs}
\usepackage{tikz}
\usepackage{textcomp}
\usepackage{appendix}
\usepackage{pifont}
\usepackage{threeparttable}

\usepackage{amssymb}
\usepackage{tabularx}
\usepackage{hyperref}
\usepackage{hypcap}
\usepackage{ulem}
\usepackage{color}

\usepackage{natbib}

\def\refone#1{{\textcolor{black}{ #1}}}    
\def\reftwo#1{{\textcolor{black}{ #1}}}    
\def\refthree#1{{\textcolor{black}{ #1}}}    

\ifCUPmtlplainloaded \else
  \checkfont{eurm10}
  \iffontfound
    \IfFileExists{upmath.sty}
      {\typeout{^^JFound AMS Euler Roman fonts on the system,
                   using the 'upmath' package.^^J}%
       \usepackage{upmath}}
      {\typeout{^^JFound AMS Euler Roman fonts on the system, but you
                   dont seem to have the}%
       \typeout{'upmath' package installed. JFM.cls can take advantage
                 of these fonts,^^Jif you use 'upmath' package.^^J}%
      }
  \else
  \fi
\fi

\ifCUPmtlplainloaded \else
  \checkfont{msam10}
  \iffontfound
    \IfFileExists{amssymb.sty}
      {\typeout{^^JFound AMS Symbol fonts on the system, using the
                'amssymb' package.^^J}%
       \usepackage{amssymb}%

      }{}
  \fi
\fi

\ifCUPmtlplainloaded \else
  \IfFileExists{amsbsy.sty}
    {\typeout{^^JFound the 'amsbsy' package on the system, using it.^^J}%
     \usepackage{amsbsy}}
    {}
\fi

\title[Experimental study of parametric subharmonic instability for internal waves]
{Experimental study of parametric subharmonic instability for internal plane waves}

\author[B. Bourget, T. Dauxois, S. Joubaud, P. Odier]%
{Baptiste BOURGET, Thierry DAUXOIS, Sylvain JOUBAUD and Philippe ODIER.}

\affiliation{Laboratoire de Physique de l'\'{E}cole Normale Sup\'{e}rieure de Lyon, 
Universit\'{e} de Lyon, CNRS, 46 All\'{e}e d'Italie, F-69364 Lyon cedex 07, France.}

\date{\today}
\begin{document}

\maketitle

\begin{abstract}
Internal waves are believed to be of primary importance as they affect ocean mixing and energy transport. Several processes can lead to the breaking of internal waves and they usually involve non linear interactions between waves. In this work, we study experimentally the parametric subharmonic instability (PSI), which provides an efficient mechanism to transfer energy from large to smaller scales. It corresponds to the destabilization of a primary plane wave and the spontaneous emission of two secondary waves, of lower frequencies and different wave vectors. Using a time-frequency analysis, we observe the time evolution of the secondary waves, thus measuring the growth rate of the instability. In addition, a Hilbert transform method allows \refone{the measurement of} the different wave vectors. We compare these measurements with theoretical predictions, and study the dependence of the instability with primary wave frequency and amplitude, revealing a possible effect of the confinement due to the finite size of the beam, on the selection of the unstable mode.
\end{abstract}

\section{Introduction}

An essential ingredient of the thermohaline circulation is the mechanism by which the denser water that was produced in the high latitude regions (colder and saltier), once it has flowed down the continental slopes towards the abyssal ocean, can come back to the surface to close the loop. This process involves an energy input to provide the gain of potential energy necessary to lift this denser water. It is believed that turbulent mixing, generated by wind and tides, is the mechanism that performs this task~\citep{Munk:DSR:66,Munk:DSR:98}. One possible way is via the breaking of internal gravity waves~\citep{Staquet:ARFM:02}, ubiquitous in the ocean, allowing a transfer of energy from large scales to small scales, where this energy is partly dissipated in heat and partly converted in potential energy through diapycnal mixing.

\refone{The detailed mechanisms for energy dissipation of internal gravity waves is still debated. Several mechanisms have been proposed:
\begin{itemize}
\item parametric subharmonic instability \citep{MacKinnonWinters,Alfordetal2007},
\item reflection on sloping boundaries \citep{DauxoisYoung},
\item scattering by mesoscale structures \citep{RainvillePinkelbis},
\item small scale bathymetry \citep{kunze2004role,Johnston,Peacock}.
\end{itemize}}
 The importance
of these four possible dissipative processes has to be estimated and compared precisely. 
It is likely that a combination of them
might be the correct answer but the usual physicists' approach, aiming at separating the different processes one by one, is 
presumably appropriate in a first stage.

Parametric subharmonic instability (PSI) is the resonant mechanism by which a primary wave is unstable to 
infinitesimal perturbations, transferring energy through the quadratic nonlinearity of the Navier-Stokes equation
to two secondary waves, satisfying temporal and spatial resonant conditions. More precisely, PSI is the class of resonant wave-wave interactions wherein energy is transferred from large scale to smaller scales and where the frequencies of the secondary waves are near half the primary frequency. Studying the particular case of a primary plane wave is very interesting since a plane wave is solution of the full inviscid nonlinear equation
for any amplitude. However, as will be discussed below, this solution becomes unstable above a given threshold, 
which can be computed analytically~\citep{McEwan1972JFM}. 

Initially, this instability was considered only for the gravity or capillary waves~(\cite{McGoldrick1965}). However, \cite{hasselmann1967} proved that this 
instability could be observed in many physical systems, and in particular for internal waves. Following this initial report, several theoretical studies have been developed for this phenomenon, deriving in particular the expression for the growth rate in presence (or absence) of viscosity~\citep{Thorpe1968,McEwan1971JFM,McEwan1972JFM,McEwan1977DAO}.  
At the same time, the first experimental observations of the instability have been reported~\citep{McEwan1971JFM,McEwan1972JFM,McEwan1977DAO,BenielliSommeria1998}. These first observations allowed \refone{the determination of} the amplitude threshold of the instability and the temporal evolution of the secondary waves amplitude. With the  \refone{development} of more powerful computers, numerical simulations have been developed on this subject~\citep{BouruetAubertot1995,Carnevale2001,Koudella2006} to estimate the energy transfer between the different scales and to determine different scaling laws. Finally, only very recently new experiments using a vertical mode-1 wave have been performed by~\cite{JoubaudPOF2012},  allowing \refone{precise measurement of} the frequencies and the wave vectors of the secondary waves and to compare theses measurements to the viscous theory. Note that a study of PSI in the very similar context of inertial waves in a rotating fluid has recently been reported by~\cite{BordesPOF2012}.

In this paper, we present  a comprehensive study of two-dimensional parametric subharmonic
instability. The results of experiments are compared with theoretical predictions.
The paper is organized as follows. 
The experimental observations for plane waves are presented in \S~\ref{observ}, followed by the presentation of the theoretical derivation of threshold, resonant conditions and growth rates in \S~\ref{theory}.
A careful comparison between experiment and theory is discussed in \S~\ref{back}. After having studied the dependence with primary wave frequency and amplitude in \S~\ref{parameters}, we present 
our conclusions and draw some perspectives .

\section{Propagation of plane waves: experimental observations}\label{observ}

\subsection{Experimental setup}\label{Experimentalsetup}

A tank, $160$~cm long and $17$~cm wide, is filled with linearly stratified salt water with constant buoyancy frequency~$N$ using the standard double bucket method.  An internal wave is generated using a wave generator similar to the one employed in previous experiments, described in~\cite{Gostiaux:EF:07} and characterized in~\cite{Mercier:JFM}. The generator is composed of stacked plates set in motion thanks to eccentered cams rotating inside the plates at constant frequency around a shaft. The motion of these plates imposes a moving boundary condition generating the desired wave. 

\begin{figure}
\centering
\includegraphics[width=1\linewidth]{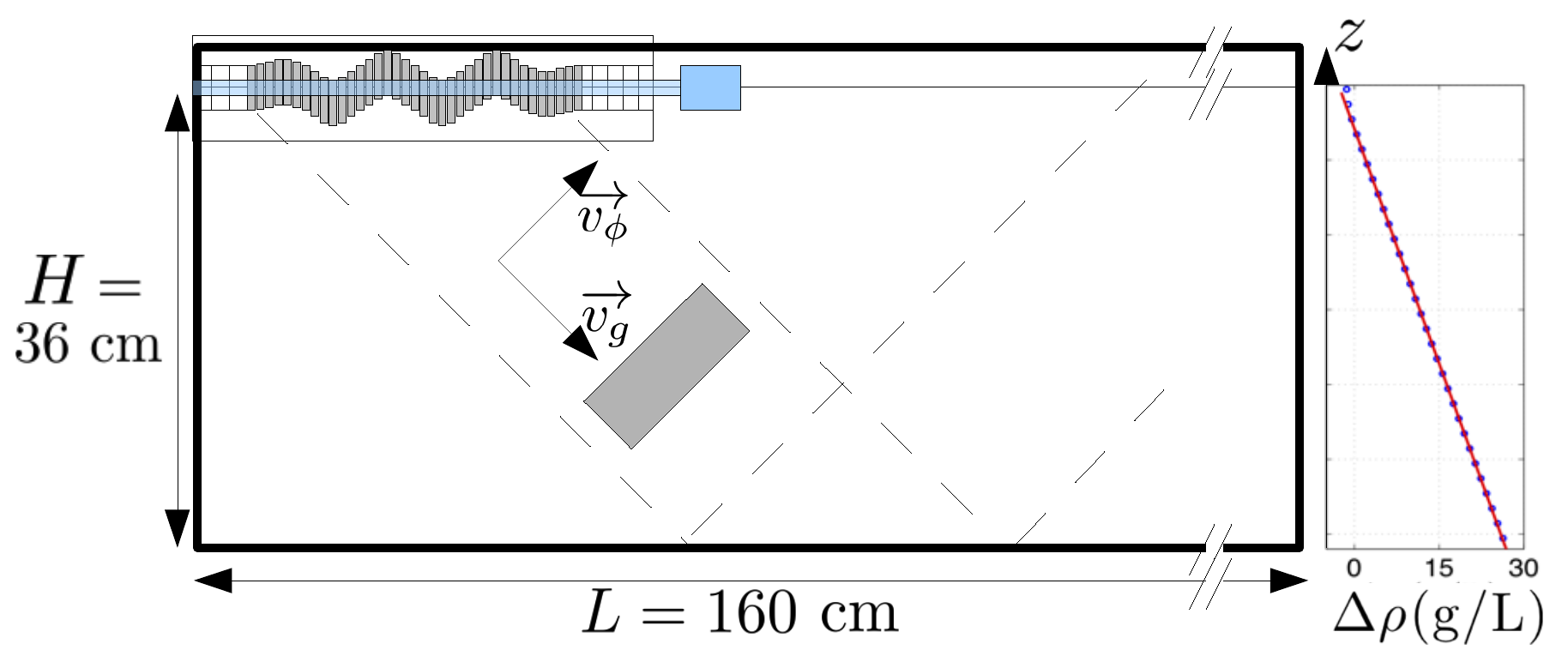}
\caption{(color online)  Sketch of the experimental setup showing the wave generator lying horizontally at the top of the wave tank. The dashed lines delimit the expected domain of propagation of the left to right propagating wave beam, once it has been emitted from the generator. For the sake of clarity, the direction of propagation of the phase and group velocities are also indicated. The tilted grey rectangle corresponds to the analysis area for the time-frequency study of section~\ref{growthrates}. The plot on the right shows the experimentally measured modification of the density due to salt, $\Delta \rho = \rho - 1000$, as a function of the water depth~$z$. Points correspond to experimental measurements while the straight line is a linear fit.}
\label{manip}
\end{figure}

The main difference with the generator described previously is that the present version allows the eccentricity of each cam to be varied. The same set of cams can therefore be used to generate different profiles, as well as identical profiles with different amplitudes. In the experiments presented here, the wave generated is a plane wave of wavelength $\lambda$=72 mm. In this case, the generator is placed horizontally as shown in Fig.~\ref{manip}. The motion of the plates is thus defined by a vertical velocity $w(x,z=0,t)=a\omega_0\cos(\omega_0 t-2\pi x/\lambda)$, $\omega_0$~being the excitation frequency, $a$ the amplitude of oscillation of the plates and $H$ the water depth. Note that to avoid spurious emission of internal waves on the extremities of the moving region, the amplitude of the plates is constant over two wavelengths in the central region, while one half-wavelength with a smooth decrease of the amplitude is added on each sides. 

The motion of the fluid is captured by the synthetic schlieren technique using a dotted image behind the tank~\citep{Dalziel:EF:00}. A camera is used to acquire images of this background at  $1.875$ frames per second. The CIVx algorithm~\citep{Fincham:EF:00}  computes the cross-correlation between the real-time and the $t=0$ background images, when the fluid is at rest. This algorithm gives the variation of the horizontal, $\tilde{\rho}_x(x,z,t)=\partial_x(\rho(x,z,t)-\rho_0(z))$, and vertical, $\tilde{\rho}_z(x,z,t)=\partial_z(\rho(x,z,t)-\rho_0(z))$,  density gradients, where
$\rho(x,z,t)$ and $\rho_0(z)$ are the instantaneous and initial fluid densities.

\subsection{Direct observation}

\begin{figure}
\centering
\includegraphics[width=1\linewidth]{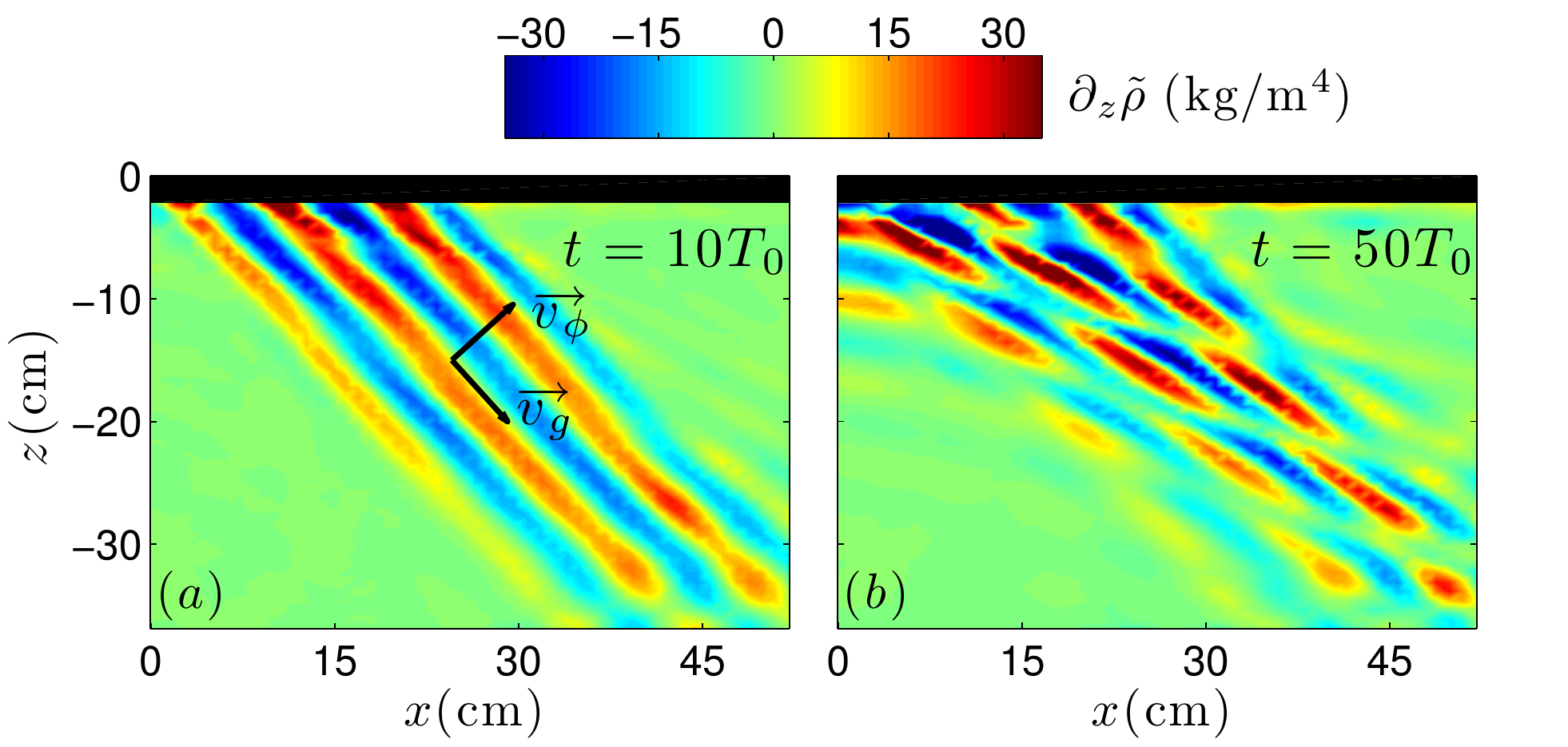}
\caption{(color online) Snapshots of the vertical density gradient field  for $t=10T_0$~(a) and  $t=50T_0$~(b) where $T_0=2\pi/\omega_0$ is the primary wave period. The wave is propagating from left to right. 
 On the left panel, the direction of the phase velocity $\protect \overrightarrow{v_{\phi}}$ and the group velocity $\protect \overrightarrow{v_{\rm g}}$ are indicated. Note that the shade scale (color scale online) is the same in both panels.
\reftwo{buoyancy} frequency is $N=0.91$~rad/s, the wave frequency is $\omega_0/N=0.74$ and the motion amplitude of the plates of the generator is $0.5$~cm. }
\label{champsdensitetemps}
\end{figure}

In an experimental configuration where the tank is filled with a linear stratification producing a \refone{buoyancy} frequency $N=0.91$ rad/s, a plane wave is generated at a frequency $\omega_0/N=0.74$, with an amplitude of the plates motion of $a=0.5$~cm. 
Figure~\ref{champsdensitetemps}(a) shows a snapshot of the density gradient field, obtained 10 oscillating periods after the wave generator was started. One can observe a plane wave extending over two wavelengths, propagating from the top left corner to the bottom right corner of the field of view, while the phase propagates from bottom left to top right. Note that the wave has reached a steady state in the visualization window. Forty oscillating periods later, a strong perturbation of the plane wave can be observed, as shown in Fig.~\ref{champsdensitetemps}(b) (see the movie in the online supplementary material). Smaller scale patterns have formed over the whole area that was initially occupied by the plane wave beam;
interestingly, these patterns extend even slightly outside of this area, in the top left corner.

\subsection{Analysis}

The measured density gradient fields are analyzed using a time-frequency representation~\citep{Flandrin:matlab:99} calculated at each spatial point
\begin{equation}
S_r(\omega,t)=\left<\left|\int_{-\infty}^{+\infty}{\rm d} u\, \tilde{\rho}_r(u)\, e^{ i \omega u}\, h(t-u)\right|^2\right>_{xz}\,,\label{Timefreqeq}
\end{equation}
where $r$ stands for $x$ or $z$ and $h$ is a smoothing Hamming window of energy unity. A large (resp. small) window provides  good frequency (resp. time) resolution. To increase the signal to noise ratio, the data is averaged on the entire area of observation. In the following, we will consider only the analysis of the vertical density gradient field, but the results are similar for the horizontal one.

\begin{figure}
\begin{center}
\includegraphics[width=0.7\linewidth]{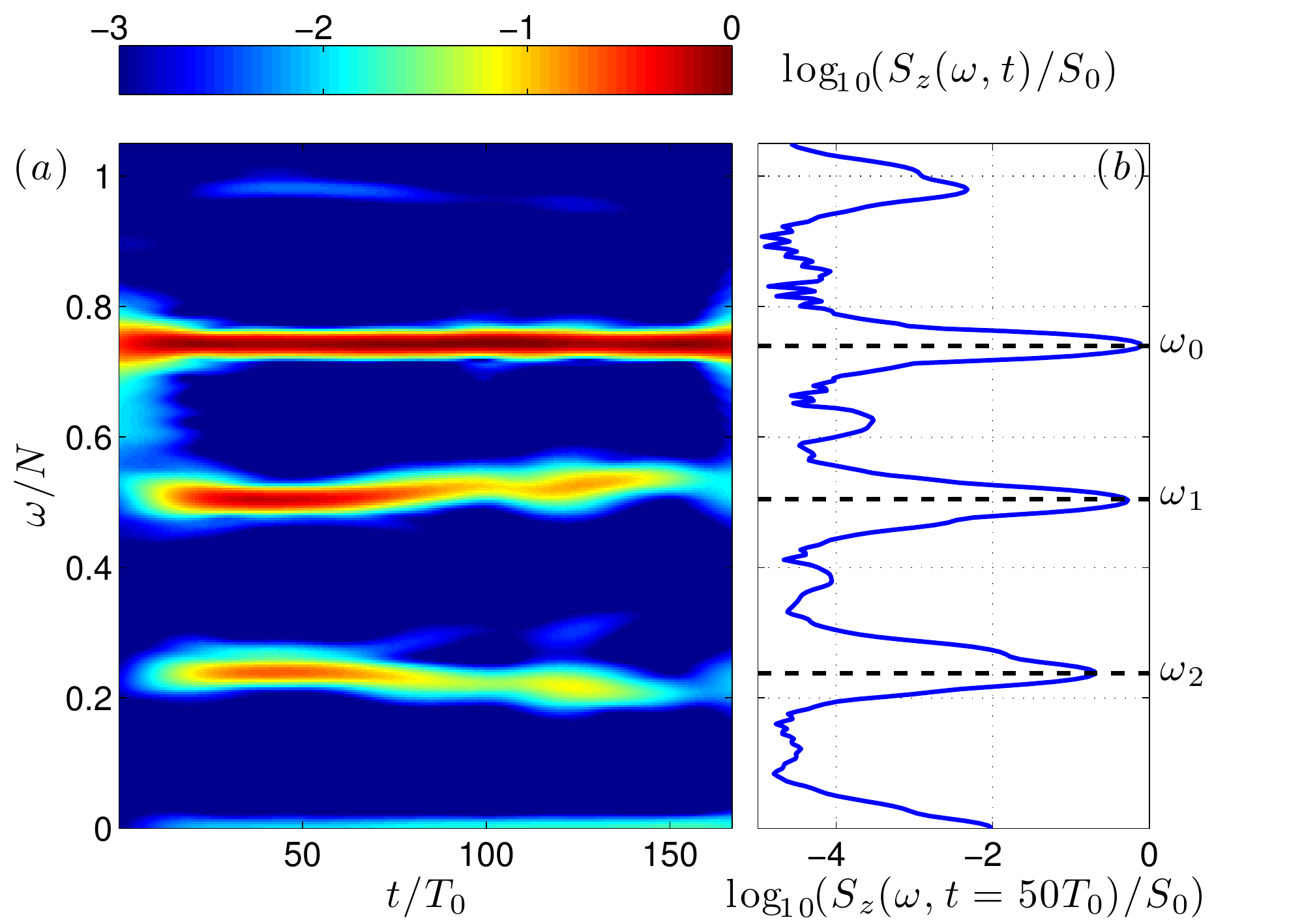}
\caption{(color online) (a) Time frequency spectrum $S_z(\omega,t)$ of the gradient density field. (b) Frequency spectrum $S_z(\omega,t=  50T_0)$.
The quantity $S_0$ is defined as the time average of the main component $S_0=\langle S_z(\omega_0,t)\rangle_t$.}
\label{tempsfrequence}
\end{center}
\end{figure}

Figure~\ref{tempsfrequence}(a) shows the time-frequency spectrum for the experiment corresponding to the snapshots presented in Fig.~\ref{champsdensitetemps}. One can clearly observe that initially, only the frequency $\omega_0/N=0.74$ is present: it corresponds to the wave produced by the generator, which we will call the {\it primary wave}. After about 10 oscillation periods, one notices the growth of two {\it secondary waves}, with the frequencies $\omega_1/N=0.50$ and $\omega_2/N=0.24$. In order to allow a better observation of these three frequencies, we present on Fig.~\ref{tempsfrequence}(b) a vertical cut of the time-frequency spectrum at time $t/T_0=50$. As can be noticed, the frequencies satisfy the condition $\omega_1+\omega_2=\omega_0$. We will come back to this feature in the following sections. It is interesting to notice that although the two secondary frequencies seem to shift slightly in the long time range, the resonance condition prevails. In addition, the amplitude of the two secondary waves seems to decrease after some time. Two other frequencies, $\omega/N=0$ and $\omega/N=0.98$, have non vanishing contributions to the signal. The first one corresponds to the mean flow generated by the plates of the wave generator   via an Archimedes' screw type of entrainment. A second possible source of mean flow could be the reflection of the primary wave at the bottom of the tank. The other additional frequency,  $\omega/N=0.98$, can be attributed to the non-linear interaction between the waves with frequencies $\omega_2/N=0.24$ and $\omega_0/N$=0.74. 
 
\begin{figure}
\includegraphics[width=1\linewidth]{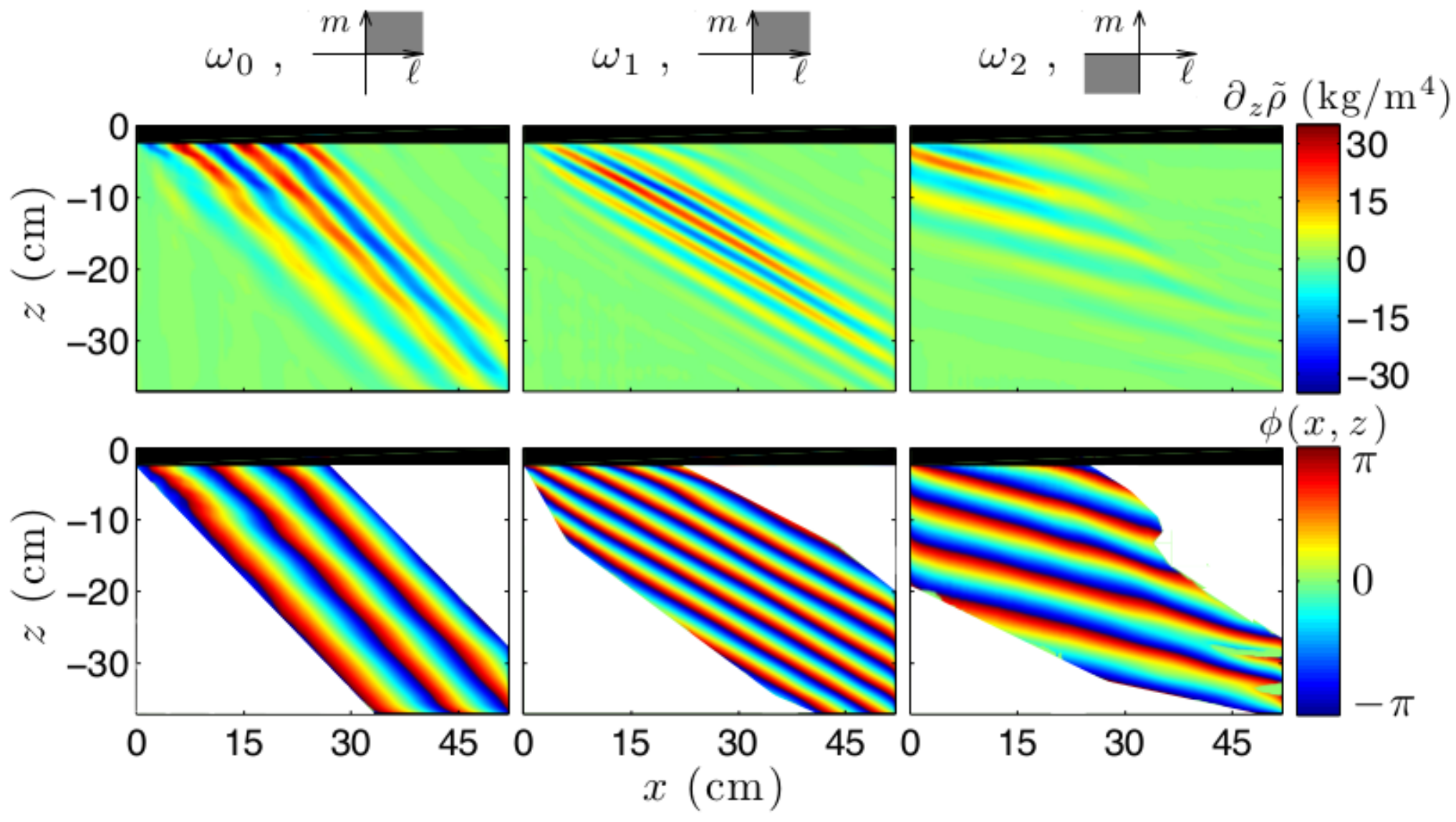}
\begin{center}
\caption{ (color online) Top row: primary (left) and two secondary waves (center and right) obtained by applying the Hilbert spatial and temporal filtering to the density gradient field at $t/T_0=50$. The sketches on top indicate the domain of wavevector $(\ell,m)$ which are kept during the spatial Hilbert filtering for each column.
Bottom row, corresponding phases associated to each frequency. The phase is displayed only
where the wave amplitude is larger than 3\% of the maximum.
The experimental parameters are $N=0.91$, $\omega_0=0.74\ N$ and $a = 0.5$~cm.}
\label{champsphase} 
\end{center}
\end{figure}

To extract more information on the various waves involved in this flow, one can filter the density gradient field around the three measured frequencies $\omega_0$, $\omega_1$ and~$\omega_2$. This filtering operation is performed using the Hilbert transform method that was developed in~\cite{Mercier:PoF:08}. This method consists in a first demodulation step with a Fourier transform in time, a time filtering around the desired frequency and then returning to real space by an inverse Fourier transform. In a second step, the signal is filtered in space, allowing \refone{separation of} the waves corresponding to the four possible propagation directions for a given frequency. The result of this operation applied to our measured density fields is shown in Fig.~\ref{champsphase}.  The \refone{filtered} density gradient fields associated with each frequency are shown in the top row, at a fixed time. Note that for the three different cases, we have kept only one quarter
of the possible wavevectors as shown in the sketches of Fig.~\ref{champsphase}. Phase and group velocities being orthogonal, the propagation direction is deduced by a rotation of 90$^\circ$, the sense of rotation being chosen to have the vertical components of the phase and group
velocities of opposite signs. \refone{For the left and center columns}, it corresponds to a propagation from the the top left corner to the bottom right one, while for the right column the wave beam goes from  the bottom right corner to the top  left one.
One can clearly observe that the filtering operation \refone{extracted} three distinct waves, each associated with its corresponding propagation angle. As mentioned, the Hilbert transform allows \refone{identification of} the orientation of the propagation of each wave, showing that the wave at frequency $\omega_2$ propagates from right to left, contrary to the two others. This explains why we observe signal outside the primary wave beam in the top left corner of Fig.~\ref{champsdensitetemps}(b), \reftwo{as mentioned in the previous section}.

This procedure also allows \refone{extraction of} the phase $\phi_i$ of the signal at a given frequency, $\phi_i(t,x,z)=\omega_{i} t\pm \ell_{i}x\pm m_{i}z$,
where $(\ell_i,m_i)$ are the horizontal and vertical components of the wavevector $\vec{k}_i$.
This is shown at $t=50~T_0$ in the bottom row of Fig.~\ref{champsphase}. Again, one can clearly observe a pattern of stripes parallel to the direction of propagation of each wave, corresponding to a phase propagating in the perpendicular direction. At a fixed time and $x$ (respectively $z$), the phase is linear with the position $z$ (resp. $x$). The components $\ell_{i}$ and $m_{i}$ of the wave vectors for each wave can then be obtained by differentiating $\phi_i(t,x,z)$ with respect to $z$ (resp. $x$). For the experiment presented in Fig.~\ref{champsdensitetemps}, one obtains $(\ell_1+\ell_2)/\ell_0 = 0.89~\pm~0.17$ and $(m_1+m_2)/m_0=0.99~\pm~0.07$. It is striking that the three vectors also satisfy a resonance condition: $\overrightarrow{k_0}=\overrightarrow{k_1}+\overrightarrow{k_2}$ \refone{within the experimental uncertainties}. Note that the larger error on the measurement of $\ell$, compared to $m$, is related to the fact that the secondary waves are rather horizontal. \refone{Consequently the horizontal projection of the wave beam, for $\omega_2$, gives us only one or two horizontal wavelengths}. It yields a poorer measurement of the horizontal component of their wavevector.

As a conclusion of this first set of experimental observations, we can say that the two secondary waves that are generated from the primary wave are the result of a resonant triad interaction. We will therefore study analytically in the next section the conditions under which such a triad interaction can develop.


\section{Parametric subharmonic instability theory}\label{theory}


\subsection{Derivation of the equations}

The two-dimensional dynamics (in the $x,z$ coordinates) of a Boussinesq fluid is usually described by the following two equations
\begin{eqnarray}
\frac{\partial b}{\partial t} + J(b, \psi) & = & -N^2 \frac{\partial \psi}{\partial x}\,,\label{Eqrho}\\
\frac{\partial \nabla^2 \psi}{\partial t} + J(\nabla^2\psi, \psi) & = & \frac{\partial b}{\partial x} +\nu \Delta^2\psi \label{Eqpsi}\,,
\end{eqnarray}
where $\psi$ is the streamfunction, while $b\equiv g \rho/\bar{\rho}$ is the buoyancy perturbation and $J$  the jacobian term defined as $J(f_1,f_2) = \partial_x f_1\partial_z f_2 - \partial_z f_1 \partial_x f_2$.  The velocity field is expressed as $\overrightarrow{v} = (-\partial_z\psi, 0, \partial_x\psi)$.

 We seek solutions of the form
\begin{eqnarray}
b &=& 
\sum_{j=0}^{2} R_j(t) e^{i (\vec{k}_j \cdot \vec{r} - \omega_j t)} +c.c.\,,\label{Sol_rho}\\
\psi &=& 
\sum_{j=0}^{2} \Psi_j(t) e^{i (\vec{k}_j \cdot \vec{r} - \omega_j t)} +c.c.\,.\label{Sol_psi}
\end{eqnarray}

Introducing these solutions into Eq.~(\ref{Eqrho}) and Eq.~(\ref{Eqpsi}), one gets
\begin{eqnarray}
\sum_{j=0}^{2} [\dot R_j - i \omega_j R_j + i N^2 \ell_j \Psi_j] e^{i (\vec{k}_j \cdot \vec{r} - \omega_j t)} + c.c. &=& -J(b, \psi)\,,\label{Eqrho2}\\
\sum_{j=0}^{2}[- \kappa_j^2 (\dot \Psi_j - i \omega_j \Psi_j) - i \ell_j R_j - \nu \kappa_j ^4 \Psi_j] e^{i (\vec{k}_j \cdot \vec{r} - \omega_j t)} +c.c. &=& - J(\nabla^2\psi, \psi)\,,\label{Eqpsi2}
\end{eqnarray}
if  $\dot R$ denotes the derivative of the amplitude $R$.

The usual inviscid linear dynamics of~(\ref{Eqrho2}) provides the polarization expression
\begin{equation}
R_j = \frac{N^2 \ell_j}{\omega_j} \Psi_j\quad \textrm{for}\quad j\,=\,0,\,1\;\textrm{or}\;2\, \label{polarization}
\end{equation}
with the dispersion relation 
\begin{equation}\omega_j=s_jN |\ell_j|/\sqrt{\ell_j^2+m_j^2} \label{dispersionrelation}
\end{equation}
where $s_j=\pm 1$ defines the sign of the wave $j$. However, the linear system is forced on the right-hand side of Eqs.~(\ref{Eqrho2}) and~(\ref{Eqpsi2}). After some \refone{manipulations}, the Jacobian terms can be written as
\begin{eqnarray}
J(b, \psi)&=&\sum_{p=0}^{2}\sum_{q\neq p}[(-\ell_{p} m_{q} + m_{p} \ell_{q}) R_{p}\Psi_{q}]e^{i [(\vec{k}_{p}+\vec{k}_{q}) \cdot \vec{r} - (\omega_{p}+\omega_{q}) t]} \nonumber\\&&\qquad- [(-\ell_{p} m_{q} + m_{p} \ell_{q})R_{p}\Psi_{q}^*]e^{i [(\vec{k}_{p}-\vec{k}_{q}) \cdot \vec{r} - (\omega_{p}-\omega_{q}) t]}  +c.c.\label{J_rho}\,,\\
J(\nabla^2\psi, \psi)&=&\sum_{p=0}^{2}\sum_{q\neq p}[(\ell_{p} m_{q} - m_{p} \ell_{q})\kappa_{p}^2\Psi_{p}\Psi_{q}]e^{i [(\vec{k}_{p}+\vec{k}_{q}) \cdot \vec{r} - (\omega_{p}+\omega_{q}) t]} \nonumber\\&&\qquad- [(\ell_{p} m_{q} - m_{p} \ell_{q})\kappa_{p}^2 \Psi_{p}\Psi_{q}^*]e^{i [(\vec{k}_{p}-\vec{k}_{q}) \cdot \vec{r} - (\omega_{p}-\omega_{q}) t]}  +c.c.\label{J_psi}\,.
\end{eqnarray}
We obtain now the evolution of a particular wavenumber component ($\vec{k}_r, \omega_r$) associated with the stream function $\Psi_r$, in which $r=0, 1$ or $2$, by averaging both the left hand side and the right hand side over the period of that wave. The resonant terms on the right hand side that will balance the left hand side will be the waves fulfilling two conditions: a spatial resonance condition
\begin{equation}
\vec{k}_0 = \vec{k}_1+\vec{k}_2 \label{spatialcondition}
\end{equation}
and a temporal resonance one
\begin{equation}
\omega_0 = \omega_1+\omega_2\,. \label{temporalcondition}
\end{equation}

Collecting resonant terms and using the polarization expression~(\ref{polarization}), the Jacobian term~(\ref{J_rho}) can then be written as
\begin{eqnarray}
J(b, \psi) &=& -(\ell_1 m_2 - m_1 \ell_2)N^2\left(\frac{\ell_1}{\omega_1}-\frac{\ell_2}{\omega_2}\right)\Psi_1\Psi_2 e^{i (\vec{k}_0 \cdot \vec{r} - \omega_0 t)} \nonumber\\
&&+(\ell_0 m_2 - m_0 \ell_2) N^2\left(\frac{\ell_0}{\omega_0}-\frac{\ell_2}{\omega_2}\right)\Psi_0\Psi_2^* e^{i (\vec{k}_1 \cdot \vec{r} - \omega_1 t)}\nonumber\\
&&+ (\ell_0 m_1 - m_0 \ell_1)N^2\left(\frac{\ell_0}{\omega_0}-\frac{\ell_1}{\omega_1}\right)\Psi_0\Psi_1^*e^{i (\vec{k}_2 \cdot \vec{r} - \omega_2 t)}\nonumber\\
&&
+ \mbox{NRT}\label{J_rho2}\,,
\end{eqnarray}
where NRT stands for non resonant terms that are not important in the problem. In the same way, one gets the Jacobian term~(\ref{J_psi})
\begin{eqnarray}
J(\nabla^2\psi, \psi) &=& \ \ (\ell_1 m_2 - m_1 \ell_2) (\kappa_1^2 - \kappa_2^2) \Psi_1 \Psi_2 e^{i (\vec{k}_0 \cdot \vec{r} - \omega_0 t)} \nonumber\\
&&-(\ell_0 m_2 - m_0 \ell_2) (\kappa_0^2 - \kappa_2^2) \Psi_0\Psi_2^* e^{i (\vec{k}_1 \cdot \vec{r} - \omega_1 t)} \nonumber\\
&&- (\ell_0 m_1 - m_0 \ell_1) (\kappa_0^2 - \kappa_1^2) \Psi_0 \Psi_1^* e^{i (\vec{k}_2 \cdot \vec{r} - \omega_2 t)} \nonumber\\
&&+ \mbox{NRT}\,.\label{J_psi2}
\end{eqnarray}
Introducing this result into equation~(\ref{Eqpsi2}), one obtains the  three following  relations between $\Psi_{r}$ and $R_{r}$ for each individual phase $\exp[{i(\vec{k}_{r} \cdot \vec{r} - \omega_{r} t)}]$ in which $r = 0, 1$~or~$2$
\begin{eqnarray}
R_0 &=&\frac{i}{\ell_0} \left[ \kappa_0^2(\dot \Psi_0 - i\omega_0\Psi_0) + \nu \kappa_0^4\Psi_0 - \gamma_0\alpha_0 \Psi_1 \Psi_2\right]\,,\\
R_{1}&=&\frac{i}{\ell_1} \left[ \kappa_1^2(\dot \Psi_1 - i\omega_1\Psi_1) + \nu \kappa_1^4\Psi_1 - \gamma_1 \alpha_1\Psi_0\Psi^*_2\right]\,,\\
R_{2}&=&\frac{i}{\ell_2} \left[ \kappa_2^2(\dot \Psi_2 - i\omega_2\Psi_2) + \nu \kappa_2^4\Psi_2 - \gamma_2 \alpha_2\Psi_0\Psi^*_1\right]\,,
\end{eqnarray}
where $\gamma_0 = +1$, $\gamma_{1,2} = - 1$ and
$
\alpha_r = (\ell_p m_q - m_p \ell_q) (\kappa_p^2 - \kappa_q^2)\,,
$ 
in which $p, q, r = 0,1,2$ or any circular permutation. 


\subsection{Slow amplitude variation}
Experimental results have strongly suggested that the amplitude of $\Psi$ varies slowly with respect to the period
of the primary wave. It is therefore appropriate to consider $\dot\Psi_j\ll i\omega_j\Psi_j$ so that differentiating~(\ref{polarization}) leads to
\begin{equation}
\dot R_j \approx \frac{\omega_j}{\ell_j}\kappa_j^2\dot \Psi_j\,.
\end{equation}
One gets for Eq.~(\ref{Eqrho2})
\begin{eqnarray}
\frac{i}{\ell_0}\left[(N^2\ell_0^2 - \omega_0^2\kappa_0^2)\Psi_0 \right.&-&\left.2 i \omega_0\kappa_0^2 \dot \Psi_0 -i \omega_0 \nu\kappa_0^4 \Psi_0 + i \omega_0 s_0 \alpha_0 (\Psi_1 \Psi_2)\right]\nonumber\\&=&(\ell_1 m_2 - m_1 \ell_2)N^2\left(\frac{\ell_1}{\omega_1}-\frac{\ell_2}{\omega_2}\right)\Psi_1\Psi_2\,,\\
\frac{i}{\ell_r}\left[(N^2\ell_r^2 - \omega_r^2\kappa_r^2)\Psi_r \right.&-&\left.2 i \omega_r\kappa_r^2\dot  \Psi_r - i \omega_r \nu\kappa_r^4 \Psi_r + i \omega_r s_r \alpha_r (\Psi_0 \Psi_p^*)\right]\nonumber\\&=&-(\ell_0 m_p - m_0 \ell_p) N^2\left(\frac{\ell_0}{\omega_0}-\frac{\ell_p}{\omega_p}\right)\Psi_0\Psi_p^*\,.
\end{eqnarray}
Each wave satisfies the dispersion relation, $N^2\ell_r^2 =\omega_r^2\kappa_r^2$. After some calculations, one gets
\begin{eqnarray}
2 i \omega_0\kappa_0^2 \dot \Psi_0=\left[i \omega_0 s_0 \alpha_0-\frac{\ell_0}{i}(\ell_1 m_2 - m_1 \ell_2)N^2\left(\frac{\ell_1}{\omega_1}-\frac{\ell_2}{\omega_2}\right)\right]\Psi_1\Psi_2-i \omega_0\nu\kappa_0^4 \Psi_0\\
= i(\ell_1 m_2 - m_1 \ell_2)\left[s_0\omega_0(\kappa_1^2 - \kappa_2^2)+\ell_0 N^2\left(\frac{\ell_1}{\omega_1}-\frac{\ell_2}{\omega_2}\right) \right]\Psi_1\Psi_2-i \omega_0\nu\kappa_0^4 \Psi_0\,,
\end{eqnarray}
which can be simplified as 
\begin{eqnarray}
\dot \Psi_0=I_0\Psi_1\Psi_2-\frac{\nu}{2}\kappa_0^2 \Psi_0.
\end{eqnarray}
Similar calculations for waves $1$ and $2$ lead to
\begin{eqnarray}
\label{equ:psi1}\dot \Psi_1=-I_1\Psi_0\Psi_2^*-\frac{\nu}{2}\kappa_1^2 \Psi_1\,,\\
\label{equ:psi2}\dot \Psi_2=-I_2\Psi_0\Psi_1^*-\frac{\nu}{2}\kappa_2^2 \Psi_2\,,
\end{eqnarray}
where
\begin{eqnarray}
I_r =\frac{\ell_p m_q - m_p \ell_q}{2\omega_r\kappa_r^2}\left[\omega_r(\kappa_p^2 - \kappa_q^2)+l_r N^2\left(\frac{\ell_p}{\omega_p}-\frac{\ell_q}{\omega_q}\right) \right].
\end{eqnarray}

\subsection{Solution}

We consider that $\Psi_0$ corresponds to the primary wave and is constant in early times since amplitudes of the secondary waves, $\Psi_1$ and $\Psi_2$ are negligible 
with respect to $\Psi_0$. One can combine equations~(\ref{equ:psi1}) and~(\ref{equ:psi2}) to get
\begin{eqnarray}
{\ddot \Psi_1} = I_{1}I_{2} \Psi_0^2 \Psi_1 - \frac{\nu^2}{4}\kappa_2^2\kappa_1^2 \Psi_1- \frac{\nu}{2}(\kappa_1^2+\kappa_2^2){\dot \Psi_1}\,. \label{eqfinal} 
\end{eqnarray}

The solution of Eq.~(\ref{eqfinal}) leads to the expression
$
\Psi_{1,2}(T)=A_{1,2}\,e^{\sigma_+ T} +B_{1,2} e^{\sigma_- T},
$ 
by introducing the growth rate
\begin{eqnarray}
\sigma_\pm = -\frac{\nu}{4}(\kappa_1^2 +
\kappa_2^2) \pm \sqrt{\frac{\nu^2}{16}(\kappa_1^2 -
\kappa_2^2)^2+I_1I_2|\Psi_0|^2}.\quad \label{Eq_lambda}
\end{eqnarray}
A vanishingly small amplitude noise induces the growth of two secondary waves. In conclusion, a primary plane wave can be unstable by a parametric subharmonic mechanism. The growth rate of the instability is not 
only function of the characteristics of the primary wave, namely its wavevector and its frequency
through the coefficients $I_1$ and $I_2$, but also its amplitude~$\Psi_0$ and the viscosity~$\nu$.


\subsection{Resonance loci and growth rates}
In the following, we consider  that only the primary wave $\Psi_0$, of
given frequency $\omega_0$, wavevector $\overrightarrow{k_0}$ = $(\ell_0,m_0)$
and  sign $s_0$, is present initially in the system, while $\Psi_{1,2}$ is experimentally only present as the noise. Without
loss of generality, we can choose $s_0=+1$ (which is given by the experimental configuration, or more precisely by the direction of rotation of the cams). The two secondary waves ($s_1$, $\omega_1$, $\overrightarrow{k_1}$) and ($s_2$, $\omega_2$, $\overrightarrow{k_2}$) which could form a resonant triad with the primary wave have to be determined using the resonance conditions (\ref{spatialcondition}) and~(\ref{temporalcondition}).
From the dispersion relation for internal waves~(\ref{dispersionrelation}),
the resonance conditions lead to
\begin{eqnarray}\label{equfinale}
s_0 \frac{|\ell_0|}{\sqrt{\ell_0^2+m_0^2}} & = & s_1 \frac{|\ell_1|}{\sqrt{\ell_1^2+m_1^2}} + s_2 \frac{|\ell_0 + \ell_1|}{\sqrt{(\ell_0+\ell_1)^2+(m_0+m_1)^2}}\,.\label{equation_k1m1}
\end{eqnarray}

\begin{figure}
\begin{center}
\includegraphics[width=0.6\linewidth]{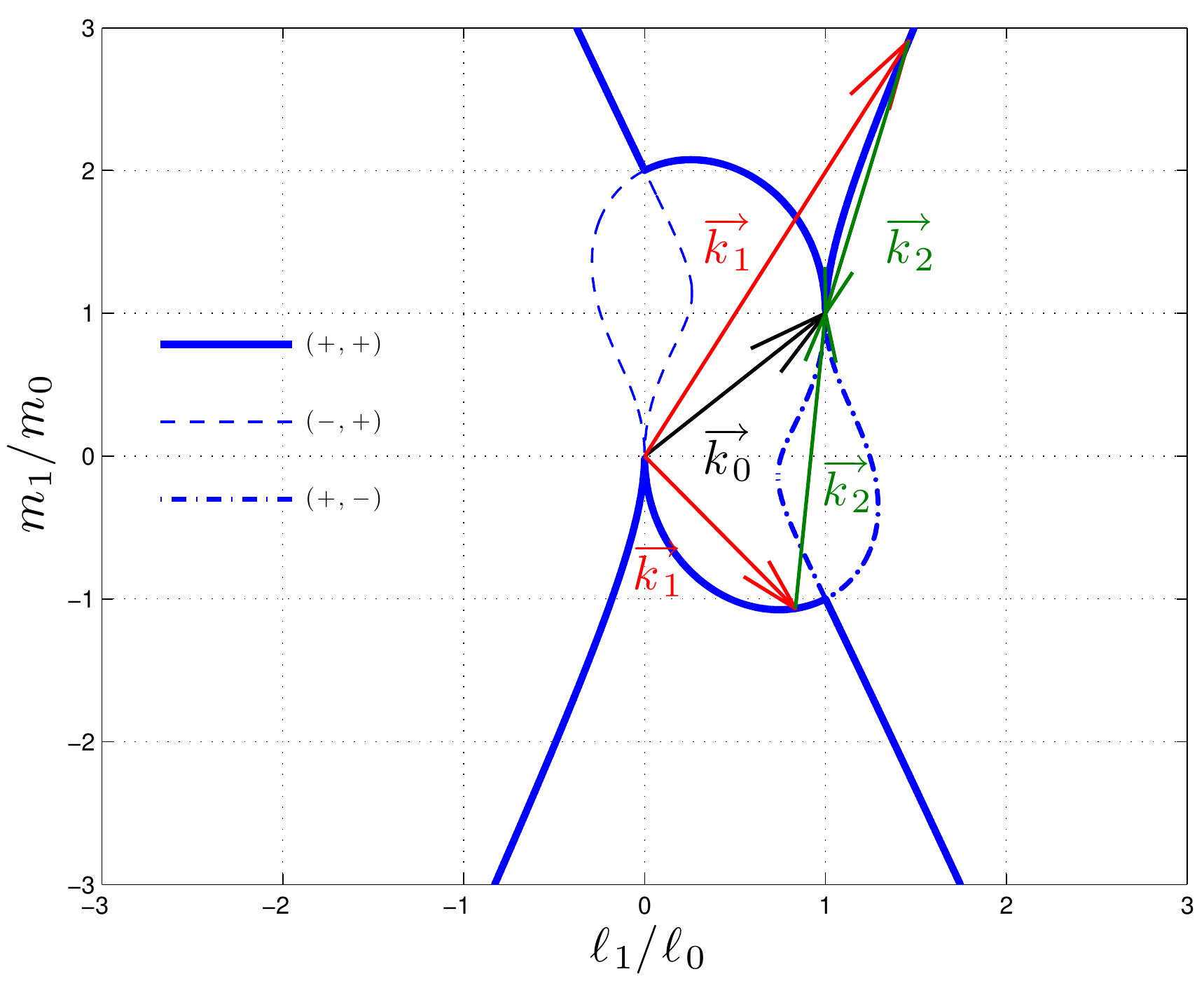} 
\caption{(color online) The curves represent the location of $(\ell_1,m_1)$ 
satisfying Eq.~(\ref{equation_k1m1}) for the three possible combinations of signs, once the wavevector $\protect \overrightarrow{k_0}=(\ell_0,m_0)$ of the
primary wave is given. The curves are represented by solid lines
when the growth rates have a positive real part, while the dashed lines corresponds to the neutrally stable cases,
for which  the real part of the growth rate is always zero in a non viscous case and negative when viscosity sets in. \refone{Two examples of vector triads ($\protect \overrightarrow{k_0}$, $\protect \overrightarrow{k_1}$,$\protect \overrightarrow{k_2}$) are shown.}}
\label{dessindelacacouete}
\end{center}
\end{figure}

For a given primary wave $(s_0, \ell_0, m_0)$,  the solution of this
equation for each sign combination $(s_0,s_1, s_2)$ is a curve in
the $(\ell_1, m_1)$ plane which is presented in Fig.~\ref{dessindelacacouete}. Since $s_0=+1$, it is then necessary to consider
four sign combinations for $(s_1,s_2)$: $(-,-)$, $(+,-)$, $(-,+)$ and
$(+,+)$. However, there is no solution of equation~(\ref{equfinale}) in the $(-,-)$ case. In addition, the combinations $(+,-)$ and $(-,+)$ are neutrally stable, {\it i.e.} the real part of $\sigma$ is always zero in a non viscous case and negative when viscosity sets in. For this reason, in what follows, we will focus only on the $(+,+)$ combination.

\refone{In Fig.~\ref{dessindelacacouete}, any point of the solid curve corresponds to the tip of the $\overrightarrow{k_1}$ vector, wich satisfy the equation \ref{equfinale} with $s_0=s_1=s_2=+1$. We obtain also $\overrightarrow{k_2}$ by construction, closing the triangle. Two examples are illustrated with solid arrows.} However, to predict which is the one expected
to be seen experimentally, one has to determine the largest growth rate. For a given sign of $m_1$, three distinct parts of the $(+,+)$ curves can be observed in Fig.~\ref{dessindelacacouete}, corresponding to $\ell_1/\ell_0>1$, $\ell_1/\ell_0<0$ and $0<\ell_1/\ell_0<1$. It will prove convenient in what follows to separate the study in these three regions, in order to evidence the direction of energy transfer (towards smaller scales or larger scales).

\begin{figure}
\begin{center}
\includegraphics[width=0.49\linewidth]{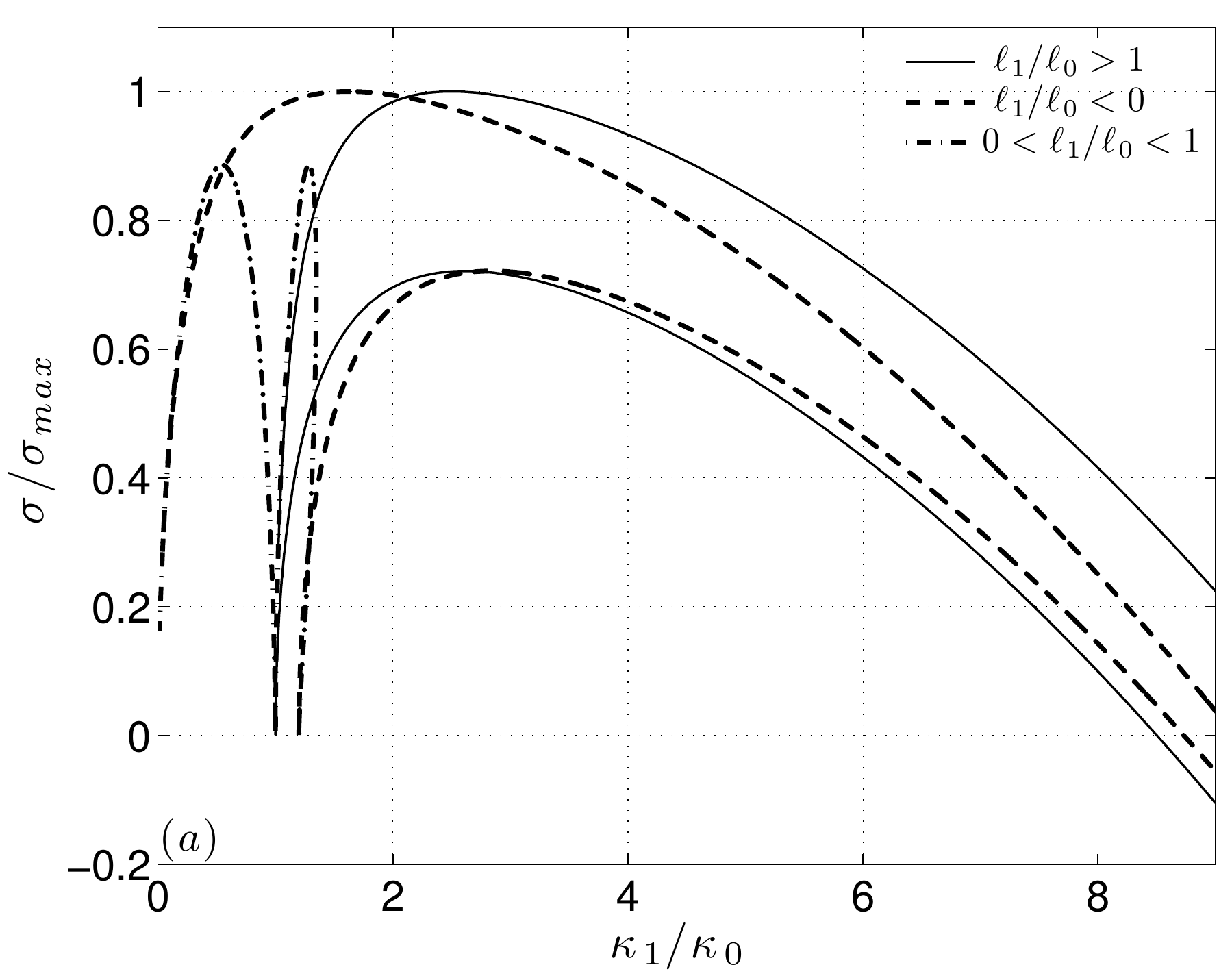}
\includegraphics[width=0.49\linewidth]{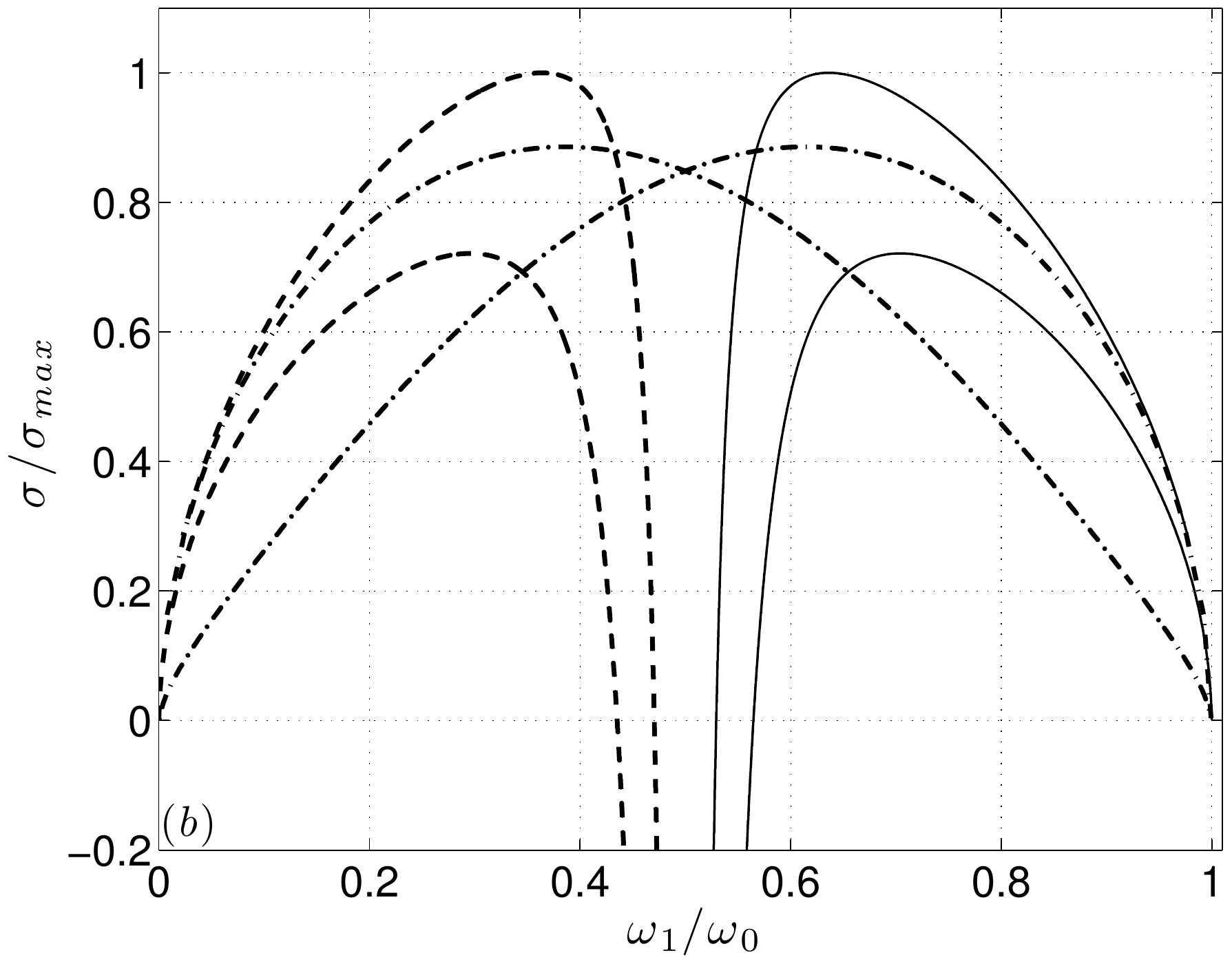}
\caption{(color online) Growth rates $\sigma$ computed from Eq.~(\ref{Eq_lambda}), as a function of the wave vector modulus~$\kappa_1$ in panel (a) and as a function of the wave frequency
$\omega_1$ in panel (b). The three possible sign
combinations have been superimposed. The primary wave vector $\vec k_0$ is chosen arbitrarily, as well as its amplitude, which value, normalized by the viscosity, is $\Psi_0/\nu=100$. $\kappa_0$ and
$\omega_0$ are respectively the wave vector modulus and the frequency of the primary wave.}
\label{growthratesenfonctiondekappaetomega}
\end{center}
\end{figure}

In Fig.~\ref{growthratesenfonctiondekappaetomega}, the predicted growth rates $\sigma$ are
plotted as a function of $\kappa_1/\kappa_0$ and $\omega_1/\omega_0$ only for the (+,+) combination, with distinct line types for the different regions defined above. There are two curves for each line type, corresponding to different signs of $m_1$ (top half and bottom half in Fig.~\ref{dessindelacacouete}).\\

One can notice that in both plots, the solid curve ($\ell_1/\ell_0>1$) and the dashed curve ($\ell_1/\ell_0<0$) reach the same maximum value. Moreover, one can observe in Fig.~\ref{growthratesenfonctiondekappaetomega}(b) that the maximum are obtained for $\omega_1=0.37\ \omega_0$ (dashed curve) and for $\omega_1=0.63\ \omega_0$ (solid curve), two frequencies which sum is equal to $\omega_0$. More generally, Fig.~\ref{growthratesenfonctiondekappaetomega}(b) shows that the solid and dashed curves are mutually symmetric with respect to the central value $\omega/\omega_0=0.5$.
It shows that if $\overrightarrow{k_1}$ is selected on the solid curve, $\overrightarrow{k_2}$ will be selected on the dashed curve and conversely. In an arbitrary way, we will always take $\overrightarrow{k_1}$ on the solid curve (largest wavenumber) and consequently $\overrightarrow{k_2}$ will be determined by the dashed curve (lowest wavenumber \refone{among the secondary waves}). 

As to the dotted dashed curve, one can notice that it has two maxima. The two secondary wave vectors are, in this case, selected in the same area ($0<\ell_1/\ell_0<1$). For all the curves, one notices that the growth rates become negative when $\kappa_1 \rightarrow \infty$ (because of viscosity).

Interestingly, the growth rate is positive for a broad range of
wavenumbers and it is rather flat on the maximum, emphasizing that
the parametric resonance is weakly selective in this regime.
The values of $\kappa_1$ corresponding to significant growth rates are of
the same order of magnitude as the primary wavenumber
$\kappa_0$, indicating that the viscosity has a
significant effect on the selection of the excited resonant triad, preventing any large wave number secondary wave to grow from the instability.
For the frequency value considered,
the maximum growth rate is obtained for $\kappa_1=2.5\  \kappa_0$ and $\kappa_2=1.62\  \kappa_0$. 
 
\section{Comparison with the experiment}\label{back}


\subsection{Wave vectors}\label{wv}

\begin{figure}
\begin{center}
\includegraphics[width=0.49\linewidth]{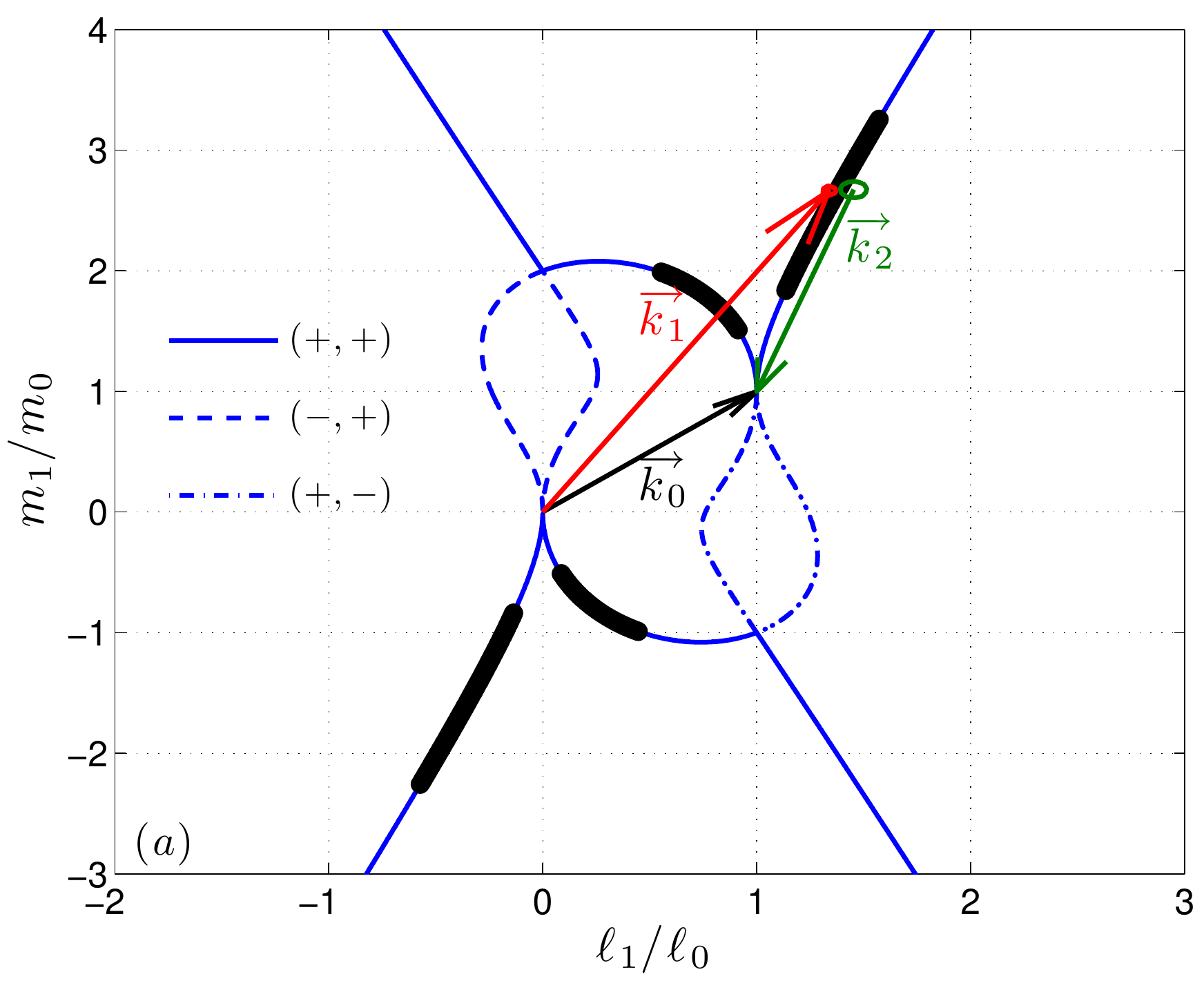}
\includegraphics[width=0.49\linewidth]{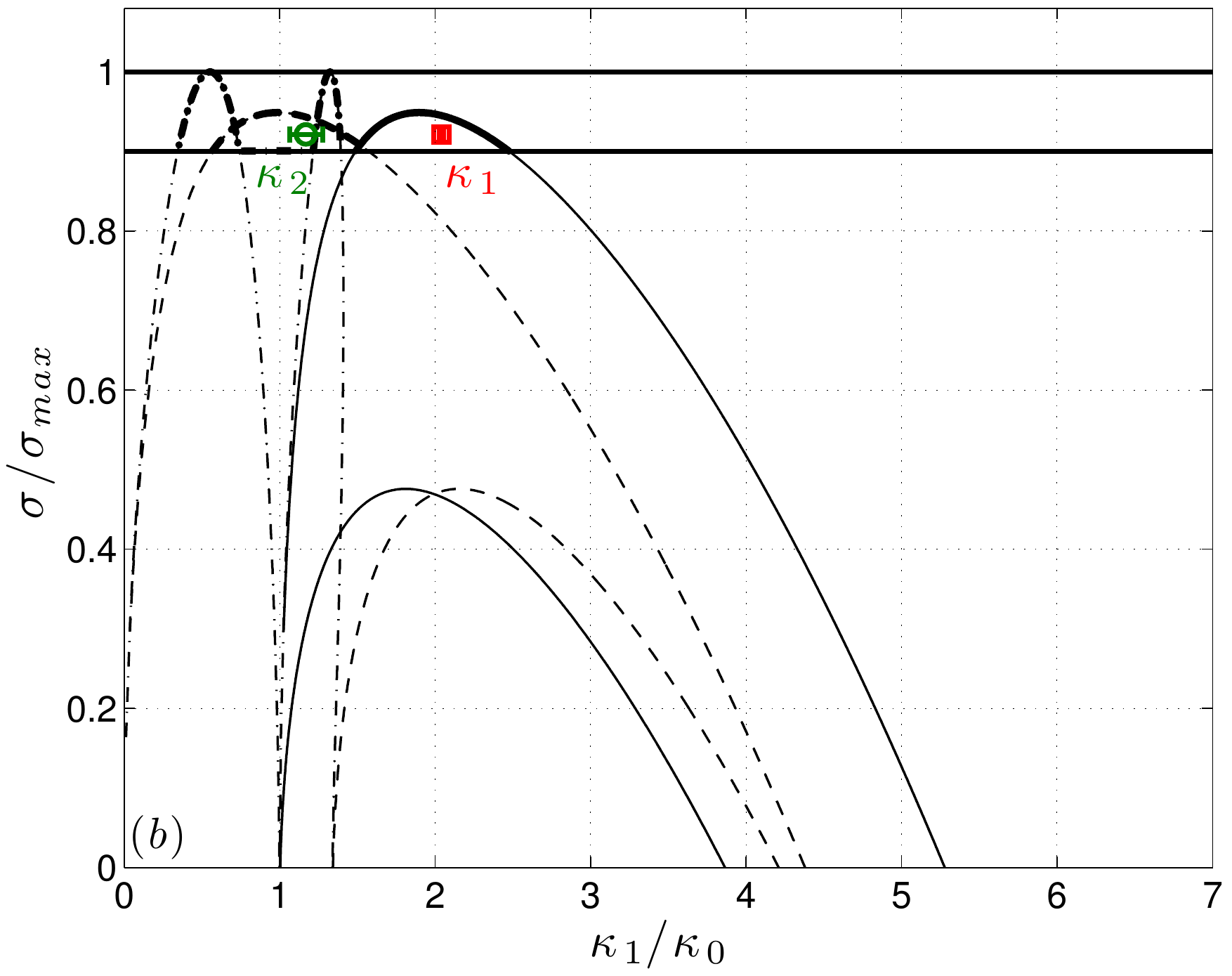}
\caption{(color online) (a) The solid blue curve represents the resonance loci for the secondary wave vector~$\protect \overrightarrow{k_1}$, in the case where the primary wave vector $\protect \overrightarrow{k_0}$ corresponds to the experiment presented in \S~\ref{observ}, {\it i.e.} $\ell_0 = 66 \pm 2$ m$^{-1}$, $m_0 = 59$ $\pm 2$ m$^{-1}$ and $\omega_0/N=0.74$. The three arrows represent the experimental measurement of the three wave vectors $\protect \overrightarrow{k_0}$ (black), $\protect \overrightarrow{k_1}$ (red) and $\protect \overrightarrow{k_2}$ (green). (b) Growth rates $\sigma$ computed from Eq.~(\ref{Eq_lambda}), as a function of the wave vector modulus, in the case where the primary wave vector $\protect \overrightarrow{k_0}$ corresponds to the experiment presented in section~\ref{observ}, as well as the primary wave amplitude ($\Psi_0/\nu=28$). The two horizontal lines mark the maximum growth rate and 90\% of the maximum. The thick lines on the left panel show the regions corresponding to a growth rate belonging to this interval. \refthree{The experimentally measured growth rate are shown for the observed wavenumber $\kappa_1$ (red square) and $\kappa_2$ (green circle). The experimental error bars on the growth rate measurement beeing large (of the order of 25\%, see text), they are not represented for these two points, for the sake of clarity.}}
 \label{wavevectors}
\end{center}
\end{figure}

In section~\ref{observ}, we observed that the experimental values of frequencies and wave vectors satisfy
 the conditions of temporal and spatial resonance~(\ref{spatialcondition}) and~(\ref{temporalcondition}). 
 The theoretical derivation gives us the dependence of the growth rate with the wave vector of one of the 
 secondary waves (the second wave is then defined by the resonance condition). We can now check 
 if the observed wave vectors correspond to the largest growth rates, as could be expected. 
 For pedagogical reasons, the value of $\Psi_0$ used to compute the plots in section~\ref{theory} 
 was arbitrarily chosen to enhance the case (external branches of the resonance loci) where the 
 energy transfer goes from large scales to smaller scales. The vector~$\overrightarrow{k_0}$ was
  also chosen arbitrarily. In this section, we will use the experimental values for~$\overrightarrow{k_0}$ 
  as well as for the amplitude $\Psi_0$. This last quantity is measured using the modulus of the Hilbert 
Transform of the density gradient field for the primary wave, calculated in the same area used 
for the measurement of the growth rate. Figure~\ref{wavevectors}(a) shows the theoretical resonance 
loci (blue curve) computed identically to the one in Fig. \ref{dessindelacacouete}, but using the 
experimental~$\overrightarrow{k_0}$. The three arrows correspond to the experimental 
measurement of the three wave vectors. As mentioned in section~\ref{observ}, it can be 
observed that the spatial resonance condition is satisfied. Figure~\ref{wavevectors}(b) 
represents the evolution of the growth rate with~$\kappa_1$. Comparing Fig.~\ref{growthratesenfonctiondekappaetomega}(a) 
and ~\ref{wavevectors}(b), is interesting to note that the reduction of the primary wave amplitude by a factor 
of 4 results in an enhancement of the ($0<\ell_1/\ell_0<1$) type of instability, with respect to the ($\ell_1/\ell_0>1$) type.

\vspace{0.5cm}

According to the theory, the secondary waves generated by the instability should 
be the ones corresponding to the largest growth rate. However, an instability is initially 
generated via a very small amplitude noise and because of the specific experimental 
conditions, not all temporal and spatial frequencies will be present in this noise with 
the same amplitude. In addition, as can be seen in Fig.~\ref{wavevectors}(b), the 
growth rate curve in the case of the ($0<\ell_1/\ell_0<1$) type of instability, although 
it has a slightly higher maximum, is much less broad than the curve for the other type, \refthree{therefore showing a very strong  wavenumber sensitivity}. 
For these reasons, we will introduce a less constrained selection criteria by allowing 
the unstable wave vectors to be associated with a growth rate belonging to an interval between $90\%$ and $100\%$ of the maximum growth rate. 
This selection criteria is illustrated in Fig.~\ref{wavevectors} (b) by thicker line and the loci for
 the corresponding wave vectors are represented in Fig.~\ref{wavevectors} (a) by thick black 
 segments. It is interesting to note that this procedure allows a selection of wave vectors in 
 different regions of the resonance loci curve. The measured wave vectors, within their 
 experimental error, fall on one pair of these selected regions, namely on the external 
 branches of the resonance loci. This selection corresponds, as mentioned earlier, to a slightly lower growth rate than the maximum, but to a broader curve.

\subsection{Growth rates}\label{growthrates}
\begin{figure}
\begin{center}
\includegraphics[width=0.6\linewidth]{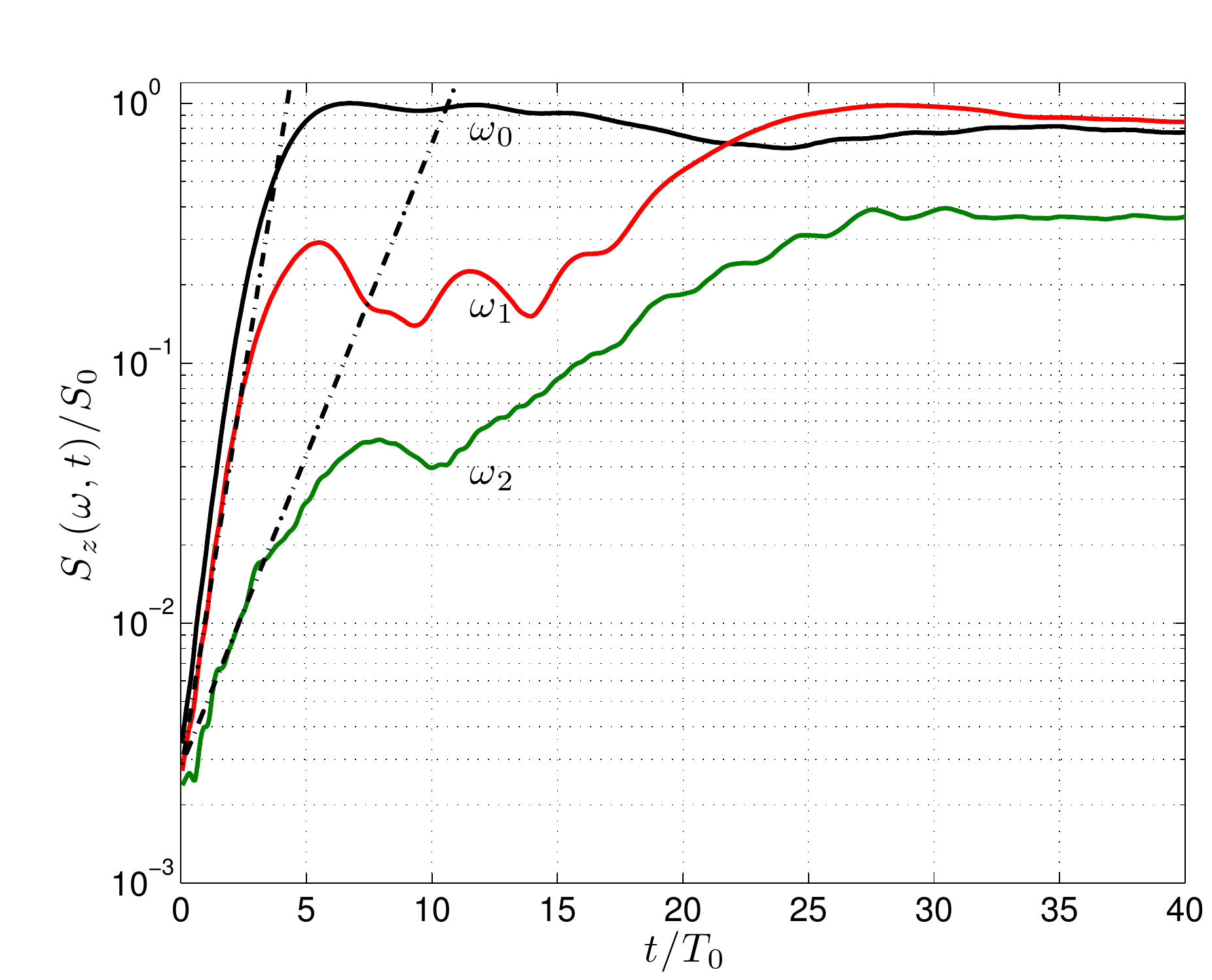}
\caption{(color online) Amplitude of time frequency representation, $S_{z}(\omega,t)$ of the secondary wave $\omega_1$ and $\omega_2$ normalized by the amplitude of the primary wave. The dashed-dotted lines show the linear fit, from which the experimental value of the growth rate $\sigma$ is extracted. The quantity $S_0$ is defined as the time average of the main component $S_0=\langle S_z(\omega_0,t)\rangle_t$.}
 \label{growthrate}
\end{center}
\end{figure}

We will now compare the experimental and theoretical values of the growth rate. The experimental determination of the growth rate $\sigma$ is obtained by calculating a time-frequency spectrum of an area taken inside the primary wave beam, as defined in Fig.~\ref{manip} by the grey area. Then, the time evolution of each partial component frequency is plotted. Figure~\ref{growthrate} shows this evolution for the same experimental run used in \S~\ref{observ}. The growth rate is then defined as the slope of the growth region. It must be noted that the amplitude of the primary plane wave is not constant on the entire beam because of the viscous damping. For this reason, the position of the area where the spectrum is computed will influence the result for the determination of $\sigma$. An estimate of the error due to this effect is therefore performed by varying this position.

\vspace{0.5cm}

For the theoretical evaluation of the growth rate, a necessary ingredient is the amplitude $\Psi_0$ of the primary wave. This amplitude cannot be obtained directly from the amplitude of oscillation of the generator plates, since there is a conversion factor between both described by~\cite{Mercier:JFM}. This factor is not well known, and strongly depends on the wave frequency, because of the change of angle it implies of the wave propagation, while the generator motion is only vertical. For this reason, $\Psi_0$ has to be measured experimentally, introducing an experimental error in the prediction. It leads to a theoretical prediction for the growth rate  $\sigma=0.087~\pm~0.005$ s$^{-1}$, while the experimental measurement gives $\sigma=0.08\pm 0.02$ s$^{-1} $. The agreement between theory and experiment is rather good.


\section{Dependence with primary wave frequency and amplitude}\label{parameters}

So far the experimental results shown correspond to a given amplitude, frequency and wave vector for the primary wave. It seems appropriate to study the effect of a variation of these parameters.

First, we have observed, based on the theoretical study, that the value of $\kappa_0$ has no effect on the selection of the secondary waves. It leaves unchanged the shape of the growth rate evolution with $\kappa_1$ and $\omega_1$ (Fig.~\ref{growthratesenfonctiondekappaetomega}). The only effect of an increase of $\kappa_0$ is an increase of the overall value of the growth rate. For this reason, it was not necessary to perform a systematic experimental study of the dependence of the growth rate with~$\kappa_0$.

\begin{figure}
\begin{center}
\includegraphics[width=0.6\linewidth]{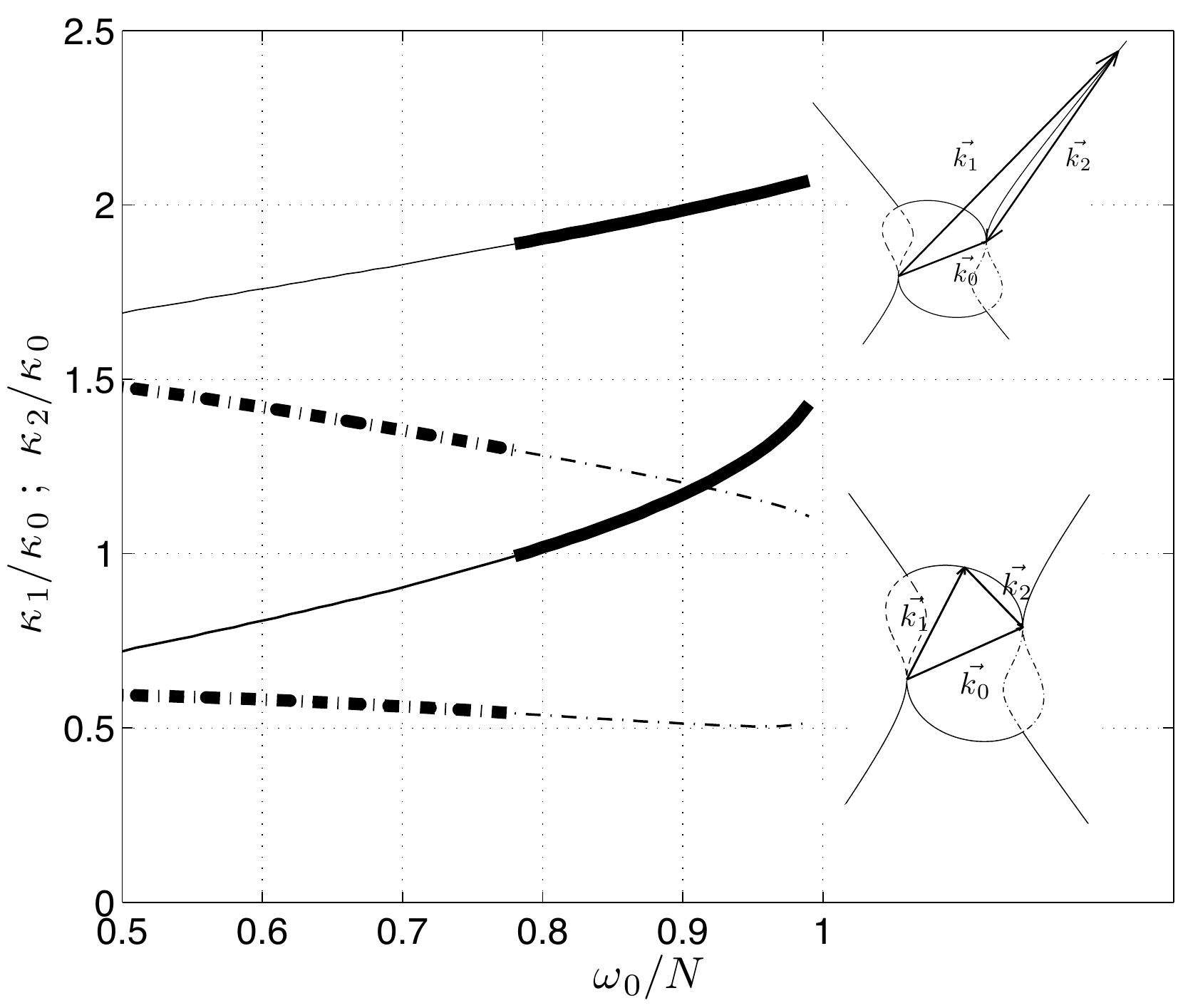}
\caption{Evolution of $\kappa_1/\kappa_0$ and $\kappa_2/\kappa_0$ with the pulsation $\omega_0$ for a fixed amplitude. Solid curve: Configuration where the triad is formed on the external branches of the resonance loci (see illustration in the top inset). Dashed dotted curve: configuration where the triad is formed on the central part of the resonance loci (see illustration in the bottom inset). The thick part of curves represent the region where the growth rate is larger in this configuration, compared to the other.}
\label{kth}
\end{center}
\end{figure}

In a first part, we will discuss the dependence with the primary wave frequency $\omega_0$. Figure~\ref{kth} shows the evolution with $\omega_0/N$ of the (normalized) norms $\kappa_1$ and $\kappa_2$ of the secondary waves corresponding to the maximum growth rate. One pair of curves (solid lines) corresponds to wave vectors on the external branches of the resonance loci and the other pair (dashed-dotted lines) to wave vectors on the central part of the resonance loci. These curves are obtained in the case where the primary wave amplitude is fixed at the value of the experimental run described in section~\ref{observ}. We can observe that in this case, when the secondary wave vectors belong to the external branches, there exist a range of frequency (for $\omega_0/N>0.78$) where the secondary wavelengths are both smaller than the primary wavelength, corresponding to an energy transfer to smaller scales. On the contrary, when the secondary wave vectors belong to the central part of the resonance loci (dashed-dotted lines), one of the secondary wavelengths is always larger than the primary wavelength, while the other is smaller, allowing for energy transfer to both smaller and larger scales. 

Out of these two sets of secondary wave vectors, the one which corresponds to the largest growth rate is illustrated on the graph by a thicker line. There is therefore a transition frequency (in this case around $\omega_0/N=0.77$) for which the instability ``jumps'' from one solution to the other.

\begin{figure}
\begin{center}
\includegraphics[width=0.7\linewidth]{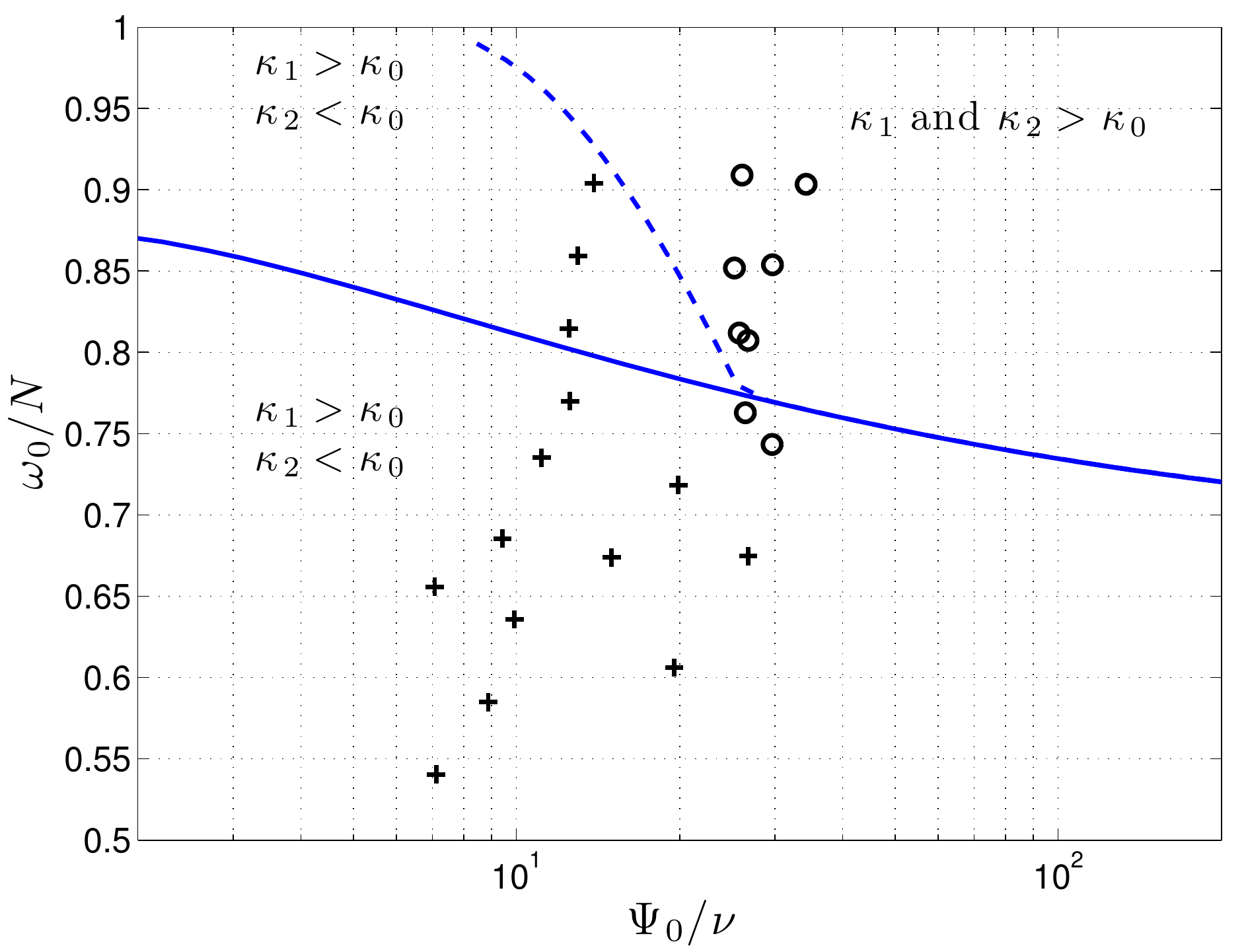}
\caption{(color online) Phase diagram showing the dependence of the PSI instability with excitation frequency and primary wave amplitude. The solid line represents the transition between the configuration where the triad forms on the external branches of the resonance loci  and the configuration where the triad is on the central part. The dashed line represents the cutoff frequency under which energy transfer to larger scales is involved in the PSI. Symbols correspond to experimental data points: ($\circ$) observed PSI, (+) no apparent PSI.}
\label{diagphase}
\end{center}
\end{figure}

As mentioned, this change in behavior is observed for a given value of the amplitude~$\Psi_0$. As this amplitude varies, the transition frequency changes. Figure~\ref{diagphase} shows (solid line) the evolution of this transition frequency as a function of the amplitude. Above this curve, the secondary waves are selected on the outer region of the resonance loci, while below the curve the secondary waves are selected in the central zone.

On the same figure, we superimposed the experimental points obtained for different values of $\omega_0$ and $\Psi_0$, showing in which cases the parametric subharmonic instability (PSI) was observed. Three different wave generator amplitudes were used to produce these series of points (cam eccentricities of $0.1$, $0.5$ and $0.75$~cm). These three series correspond to the three almost vertical series of black symbols on the graph. One must keep in mind that the wave amplitude is computed from the measured signals, therefore a fixed amplitude of the wave generator does not correspond exactly to a fixed wave amplitude.

\vspace{0.5cm}

It can be noted that for large enough excitation amplitudes, PSI is observed, but mostly above the transition line. Actually, after careful observation, it turns out that the two points with PSI below this line are also case of PSI with wave vectors on the external branches of the resonance loci. This feature may be related to the uncertainty on the frequency selection mentioned in section \ref{wv}. 

Our guess is that PSI cannot develop in the present experimental tank when one of the secondary waves has a larger wavelength than the primary wave. No cascade to larger scale is allowed. The reason for this behavior could be the too small number of wavelengths which is excited by the generator.

For the smallest excitation amplitude, no PSI was observed during the recording, \refthree{even for experimental runs located above the solid line of figure~\ref{diagphase}}. Equation~(\ref{Eq_lambda}) shows that unstable solutions are possible only if the amplitude of the primary wave exceeds a viscous threshold value given by $2\sqrt{\nu\kappa_1\kappa_2/(I_1I_2)}$. However, the excitation amplitude is one or two orders of magnitude larger than this viscous threshold which cannot explain the absence of the instability. This behavior could again be related to the size of the expected wavelengths at this amplitude. Indeed, it can be observed in Fig.~\ref{kth} that the lower solid line crosses the ordinate 1 at a given frequency. \refthree{This implies that}, below this frequency, \refthree{although the secondary waves are selected on the external branches,} a transfer to larger scales is \refthree{however }involved. We have computed theoretically, as a function of the wave amplitude, the  threshold frequency under which \refthree{a transfer to larger scales is present for secondary waves selected on the external branches.} It is plotted in Fig.~\ref{diagphase} as a dashed line. The small amplitude series, where no PSI is observed, happens to be on the left of this line, therefore in the region involving transfer to larger scales. As in the case of small frequencies, this might explain the absence of observed PSI. \refthree{To summarize, the only region in the phase space where no energy is transfered to larger scales is the top right corner (above the solid line and right of the dashed line) and it is the only region where we observe PSI, which seems to agree with our assumption of limitation due to the finite size of the primary wave beam.}


\section{Discussion}

Using a synthetic schlieren technique in a linearly stratified fluid, we have observed the production, from a primary internal plane wave, of two secondary waves of smaller frequency and wavelength. The mechanism at play is a 3-wave resonant interaction, satisfying a temporal resonance condition for the three pulsations $\omega_0=\omega_1+\omega_2$ and a spatial resonance condition for the three wave vectors,  $\overrightarrow{k_0}=\overrightarrow{k_1}+\overrightarrow{k_2}$. This resonant interaction can be developed analytically, in order to produce theoretical predictions that we compare to our measurements, which are derived from applying a time-frequency spectrum and a Hilbert transform to our data. A good agreement was found for the wave vectors of the secondary waves, as well as for the instability growth rate. 

The relevance of PSI for oceanic internal waves has been recently renewed by several numerical simulations with realistic conditions (see for example \cite{Hibiyaetal} or \cite{MacKinnonWinters}), which have been followed 
by very interesting field measurements which also pointed to the presence of PSI activity in the ocean. One might in particular refer to measurements close to Hawaii by \cite{RainvillePinkel}, \cite{CarterGregg} or \cite{Alfordetal2007}.

\begin{figure}
\begin{center}
\includegraphics[width=0.49\linewidth]{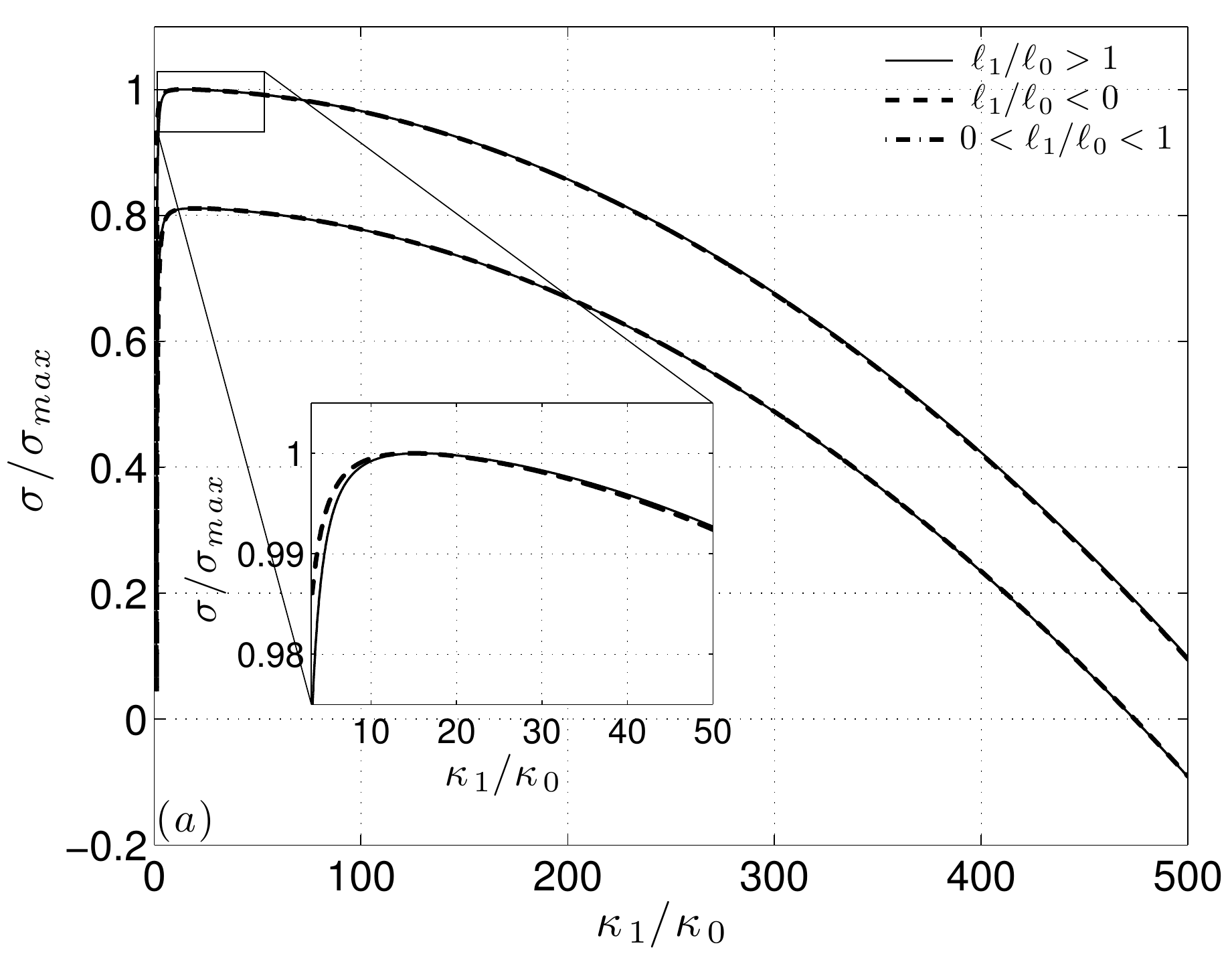}
\includegraphics[width=0.49\linewidth]{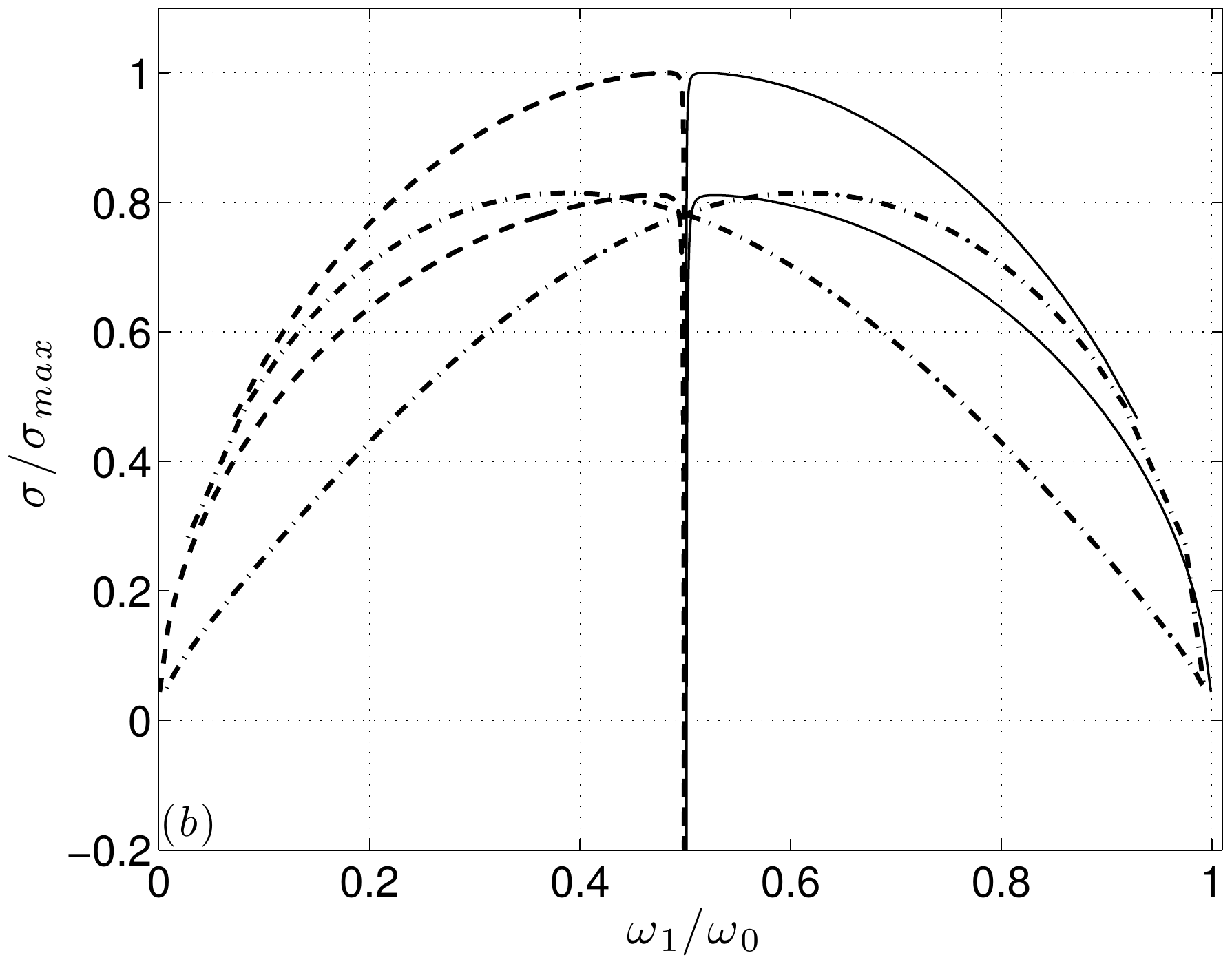}
\caption{(color online)  Growth rates $\sigma$ as a function of the wave vector modulus~$\kappa_1$ in panel (a) and as a function of the wave frequency
$\omega_1$ in panel (b) for typical oceanic parameters (see text).}
\label{PSI_ocean}
\end{center}
\end{figure}

{However, in the ocean, due to the large spatial scales, the role of viscosity is reduced compared to the experiment. In order to illustrate this point, let us consider a typical oceanic internal wave, measured in situ by observing the periodic oscillations of isotherms (see figure 1 in~\cite{CairnsWilliams} or figure 6.13 in~\cite{Sutherland:10}). Theses oscillations have an amplitude of about 20 m, with a period of about half a day. 
Ocean internal waves typically have wavelengths from hundreds of meters to tens of kilometers \citep{GarrettMunk,LeBlondMysak,Susantoetal}: as a typical value, we shall take 1000 m. 
{Compared to our experiment, it yields an amplitude of the stream function $\Psi_0$$/\nu$ about four orders of magnitude larger. We therefore recomputed as an example (see Fig.~\ref{PSI_ocean}) the curves of Fig.~\ref{wavevectors}(b), as well as the corresponding plots as a function of the wave frequency, for an amplitude 10000 times larger than the experiment. As can be observed in Fig.~\ref{PSI_ocean}(a), the most unstable mode corresponds to the situation where the norm of both secondary wave vectors are much larger (10 to 20 times in our example) than the norm of the primary wave vector, leading to ${\overrightarrow{k_1}\simeq\overrightarrow{k_2}}$ and correspondingly (Fig.~\ref{PSI_ocean}(b)) to $\omega_1\simeq\omega_2\simeq\omega_0/2$. This is what led historically to call this resonant interaction "Parametric subharmonic instability'', because of the similarity with parametric instability, like that occurring for a pendulum with oscillating suspension. }
}

In our experimental tank, with dimensions orders of magnitude smaller than in the ocean, we observe that viscosity has a {much larger} effect on the selection of the unstable modes. Indeed, because of viscosity, $\kappa_1$ and $\kappa_2$ are of the same order of magnitude as $\kappa_0$. In addition, a second effect of viscosity is to allow two different behaviors for the instability. First, the instability can generate two new internal waves whose wavelengths are smaller than the primary wavelength. We have then an energetic transfer to smaller scales, as is the case in the ocean. But theoretically, when $\Psi_0/\nu$ is small enough, the instability can also generate two secondary internal waves with one of the wavelengths larger than the primary wavelength, and the other one smaller. In this case, the energetic transfer will be to larger and smaller scales simultaneously. However, we never observed experimentally this second behavior, although it was expected from the analytical study. We observe the instability, on a plane wave, only when the theory predicts two smaller wavelengths. It is possible that the finite size of the beam impedes the development of the secondary wave, which has a larger wavelength. It could be interesting to perform some numerical simulations, with a confined beam to confirm this behavior.

Another issue that needs to be discussed is the physical location of the birth of the secondary waves. In all the experiments performed, the instability appears first in a region very close to the wave generator. This is understandable because it is the location where the primary wave first appears and where its amplitude is maximum. Then it occupies the whole volume. 

In addition to the importance of the mechanism for the dissipation of waves, an effect on the background
media in which waves are propagating can be expected: indeed, the primary wave could give rise through this 
instability to two secondary waves which are above the overturn threshold. Because of the latter, the stratification will
evolve through these mixing events, influencing through this feedback effect the threshold for the mechanism itself. 

\section*{Acknowledgements}
We thank  G. Bordes, H. Scolan, C. Staquet for insightful discussions. This work has been partially supported by the ONLITUR grant (ANR-2011-BS04-006-01). This work has been partially achieved thanks to the resources of PSMN (P\^ole Scientifique de Mod\'elisation Num\'erique) de l'ENS de Lyon.

\section*{Supplementary material}
Supplementary materials are available at journals.cambridge.org/flm. 

%

\end{document}